%% file: Draft.tex
\documentclass[draft]{agujournal2019}
\usepackage{url} 
\usepackage{lineno}
\usepackage{soul}
\usepackage{physics}
\usepackage{units}
\usepackage{cleveref}
\usepackage{amssymb}
\usepackage{tabularx}
\usepackage{listings}
\usepackage{enumitem}
\usepackage{booktabs}
\usepackage{cancel}
\usepackage{array,etoolbox}
\preto\tabular{\setcounter{magicrownumbers}{0}}
\newcounter{magicrownumbers}

\draftfalse

\input{shortcuts.tex}

\newcommand{\krbar}{\ensuremath{\bar{k}_r}}
\newcommand{\ugbar}{\ensuremath{\bar{u}_g}}
\newcommand{\tauacc}{\ensuremath{\tau_{\text{acc}}}}
\newcommand{\taused}{\ensuremath{\tau_{\text{sed}}}}
\newcommand{\taustar}{\ensuremath{\tau^*}}
\newcommand{\tauf}{\ensuremath{\tau_f}}
\newcommand{\taupeak}{\ensuremath{\tau_\mathrm{peak}}}
\newcommand{\deltai}{\ensuremath{\Delta_i}}

\newcommand{\wf}{\ensuremath{w_f}}
\newcommand{\rhow}{\ensuremath{\rho_w}}

\newcommand{\Mo}{\ensuremath{M_0}}
\newcommand{\Mc}{\ensuremath{M_c}}
\newcommand{\Mr}{\ensuremath{M_r}}
\newcommand{\W}{\ensuremath{W}}

\newcommand{\Mcdot}{\ensuremath{\dot{M}_c}}
\newcommand{\Mrdot}{\ensuremath{\dot{M}_r}}
\newcommand{\mc}{\ensuremath{m_c}}
\newcommand{\mr}{\ensuremath{m_r}}

\newcommand{\LWP}{\ensuremath{\text{LWP}}}
\newcommand{\RRav}{\ensuremath{\langle RR \rangle}}

\begin{document}

\title{Single-parameter effective dynamics of warm cloud precipitation}

\authors{Shai Kapon\affil{1}, Nadir Jeevanjee\affil{2}, Anna Frishman\affil{1} }

\affiliation{1}{Department of Physics, Technion Israel Institute of Technology}
\affiliation{2}{Geophysical Fluid Dynamics Laboratory, Princeton, NJ, USA.}

\affiliation{1}{32000 Haifa, Israel}
\affiliation{2}{201 Forrestal Rd, Princeton, NJ 08540, USA}

\correspondingauthor{Anna Frishman}{frishman@technion.ac.il}
\correspondingauthor{Nadir Jeevanjee}{nadir.jeevanjee@noaa.gov}

\begin{keypoints}
\item Analytical bulk approximations are developed for spectral bin simulations of warm rain microphysics
\item Cloud observables collapse as a function of a single non-dimensional parameter $\mu$, the ratio of accretion and sedimentation time scales.
\item A constant bulk fall speed performs surprisingly well in capturing bulk dynamics of sedimentation
\end{keypoints}

\begin{abstract}
Cloud observables such as precipitation efficiency and cloud lifetime are key quantities in weather and climate, but understanding their quantitative connection to initial conditions such as initial cloud water mass or droplet size remains challenging. Here we study the evolution of cloud droplets with a bin microphysics scheme, modeling both gravitational coagulation as well as fallout, and develop analytical formulae to describe the evolution of bulk cloud and rain water. We separate the dynamics into a mass-conserving and fallout-dominated regime, which reveals that the overall dynamics are governed by a single non-dimensional parameter $\mu$,  the ratio of accretion and sedimentation time scales. Cloud observables from the simulations accordingly collapse as a function of $\mu$. We also find an unexpected relationship between cloud water and accumulated rain, and that fallout can be modeled with a bulk fall speed which is constant in time despite an evolving raindrop distribution.
\end{abstract}


\section{Introduction}
Cloud microphysical processes, such as the formation of cloud and rain drops, are a critical component of the hydrological cycle and hence of Earth's weather and climate. But, these processes exhibit variability over a wide range of scales, from microns to meters, which moreover are  far smaller than that of  large-scale weather and climate processes. As such, the modeling of cloud microphysics is both critically important and notoriously difficult. These difficulties are brought to the fore by state-of-the-art global storm-resolving models which explicitly simulate convection but must still rely on uncertain microphysics parameterizations \cite{sullivan2021,atlas2024,naumann2024}, as well as by proposals for geoengineering such as marine cloud brightening and cirrus cloud thinning which rely heavily on uncertain microphysical processes. \cite{lohmann2017,feingold2024}.

This state of affairs has motivated promising new approaches to microphysical parameterization \cite<e.g.>{morrison2015a,dejong2022,hu2024}. At the same time, however, it also seems worthwhile to step back and deepen our physical intuition for microphysical processes, to strengthen the foundations of the field and to help guide its ongoing development. In this spirit, we will focus here on developing an analytical understanding of a small set of microphysical processes. We will focus on warm-rain processes, taking advantage of their relative simplicity and reduced uncertainty as compared to ice processes. As a first step, we will consider these microphysical processes in isolation from the clouds environment. As the initial formation of rain drops (`autoconversion') can be significantly impacted by the environment~\cite{Stevens_understanding_2008,Seifert_microphysical_2010}, our main focus will be the regime where some rain drops have already formed. Drop evolution is then dominated by accretion of cloud drops onto rain drops and the fallout of rain drops. These latter processes are key determinants of certain cloud observables of interest, including precipitation efficiency, cloud lifetime, and rain rates. Note that precipitation efficiency in particular has recently been highlighted as a key determinant of climate and climate sensitivity \cite{zhao2014,zhao2016a,li2022,lutsko2023}. 

Our approach will be hierarchical, in the spirit of climate model hierarchies employed elsewhere in climate science \cite{held2005,jeevanjee2017a}, and building on classical coarse-graining approaches in cloud microphysics~\cite{Kessler_on_1969,berry1974analysis,seifert_double-moment_2001,khain_representation_2015}. The highest rung of our hierarchy  will be a ``spectral bin" scheme which resolves the evolution of a droplet population as a function of droplet size~\cite{khain_representation_2015}. We will then develop a much simplified `bulk approximation' to these dynamics in terms of only two variables, cloud water mass and rain water mass as first suggested in~\cite{Kessler_on_1969}, which reasonably emulates the behavior of the bin scheme in our set-up. The bulk approximation will allows us to identify a non-dimensional parameter $\mu$ which governs the dynamics, and will also reveal a surprising relationship between cloud water mass and accumulated rain. Finally, the effective parameter $\mu$ and the solutions to the bulk equations will shed light on the lowest rung of our hierarchy --- the cloud observables identified above, providing insights into the organizational principles determining them.


\section{Simulations: formulation and characterizations} \label{sec_sims}

\subsection{Definition of the problem}
Our approach will be to model a static, non-ascending warm cloud of fixed depth $L$ which does not mix with its environment. We will thus consider an already-formed droplet population with no condensation source of additional water mass, no inhomogeneities from turbulence, no variations with height (averaging over the entire cloud column), and no dilution or evaporation from entrainment. The only physics we consider is gravitational, i.e. the collision and coalescence of drops due to nonzero relative fall speeds (i.e. "gravitational coagulation"), as well as the removal of larger drops by fallout (rain).  These are strong assumptions, but such a restriction in scope seems to be a necessary first step in obtaining the kind of understanding motivated in the introduction, and   has some precedent in the literature~\cite{seifert_double-moment_2001,falkovich2006rain}. 

Given this idealized static cloud, we model its droplet population using the bin microphysics approach, where the population is represented by a distribution $n(\nu,t)$. Here, $n(\nu,t)\dd\nu=n(r,t)\dd r$ gives the number of drops with volumes $\nu\in[\nu,\nu+\dd\nu]$ (or radii $r\in[r,r+\dd r]$) in a unit volume of the cloud~\cite{yau1996short,morrison2020confronting}. We also denote $m(r)=\rhow \nu(r) n(\nu(r))\dd r$ as the mass in radius $r$ per unit volume, and $\dd m(r)/\dd \log(r)$ the mass distribution (for a logarithmic grid), where $\rhow=10^3\ \kg/\meter^3$ is the density of water. Since our focus is on collision-coalescence and fallout,  we describe the time evolution of $n(\nu)$  by the Smoluchowski coagulation equation along with a fallout sink, following \citeA{falkovich2006rain}: 
\begin{linenomath}
\begin{equation}
    \label{eq : Smoluchowski in volume}
    \begin{split}
      &\pdv{}{t}n(\nu)=-\frac{u_g(\nu)}L n+\frac{1}{2}\intop_{0}^{\nu}H(\delta,u)n(\delta)n(u)\dd u-\intop_{0}^{\infty}H(V,\nu)n(V)n(\nu)\dd V\\
      &\delta=\nu-u \ .
        \end{split}
\end{equation}
\end{linenomath}
The first term on the right-hand-side is obtained from an advection term $-(\vec{v}\cdot\grad)n$ assuming $\vec{v}=-u_g(\nu)\hat{z}$ where $u_g(\nu)$ is the gravitational sedimentation velocity of a drop of volume $\nu$, and we have averaged over the cloud depth  $L$. The two other terms represent, respectively, the increase in the number of drops with volume $\nu$ due to coalescence of two other smaller drops, and the decrease in drops with volume $\nu$ due to a coalescence event with a drop of any other size. Here $H(V,\nu)$ is the coagulation kernel corresponding to the probability of two drops with volumes $\nu$ and $V$ to collide in unit time. As mentioned above we consider the gravitational kernel, which takes the form 
\begin{linenomath}
\begin{align}
\label{def : gravitational kernel}
H(V,\nu)=K\left[\left(\frac{3V}{4\pi}\right)^{\nicefrac{1}{3}},\left(\frac{3\nu}{4\pi}\right)^{\nicefrac{1}{3}}\right]; && K(R,r)&=\pi(R+r)^{2}\left|u_{g}(R)-u_{g}(r)\right|E(R,r)
\end{align}
\end{linenomath}
representing the cross section for collision between two drops, each falling with its own terminal speed~\cite{yau1996short}. The function $E(R,r)$ is the so-called collision efficiency, taking into account aerodynamic effects. Note that here we do not include the effect of turbulence on the collision rate, which is significant for collisions between drops of comparable size and small drops (e.g. $< 40\mu m$ ), important at the early stages of the drop population development~\cite{wang2009role,franklin2007statistics,khain2007critical,grabowski2013growth}. Instead, we focus on a regime where some larger drops have already formed (i.e. after the so-called condensation-coalescence bottle-neck is crossed~\cite{grabowski2013growth}, and the dynamics become dominated by processes involving drops of different sizes, in particular larger drops ($\gtrsim 30\mu m$)). This partially justifies our neglect of turbulence.

We are interested in characterizing the gross observables of this system (accumulated rain, lifetime, and rain rate) as a function of its initial bulk properties. In addition to the fixed cloud depth $L$, we follow \citeA{seifert_double-moment_2001} and take initial conditions $(M_0,N_0,r_0)$  where $M_0$ is the initial liquid water content (LWC) in mass per unit volume, $N_0$ is the initial number of drops per unit volume, and $r_0$ the initial characteristic radius of droplets,  defined to be the maximum of $n(r)r^3$, i.e. the radius of maximum mass. As is standard for the collision-coalescence stage of the dynamics~\cite{berry1974analysis,seifert_double-moment_2001}, the initial distribution function for all of our simulations is taken to be a Gamma function, 
\begin{linenomath}
\begin{equation}
    n(\nu, t=0)=A(\rhow\nu)^{p}\exp[-B\rho\nu] \ .
\end{equation}
\end{linenomath}
The three parameters $(A,B,p)$ appearing in the distribution can be directly related to the three physical bulk parameters $(M_{0},N_{0},r_{0})$ discussed above. 

\subsection{Simulation details and general characteristics}
We solve \eqnref{eq : Smoluchowski in volume} following the spectral bin algorithm introduced in \citeA{falkovich2006rain}. The droplet size distribution, $n(\nu,t)$, is represented as the set of concentrations $n_{i}(t)$ of droplets with volumes $\nu_{i}$ where $\nu_{i}=\frac{4}{3}\pi r_{i}^{3}$ and $r_{i}$ is part of a discretized drop radii space. For the collision efficiency $E(R,r)$ we use the values given in~\citeA{pinsky2001collision}, and we take the values for the sedimentation velocity $u_g(\nu)$ given in \citeA{seifert_double-moment_2001}. The grid of drop radii is taken to be logarithmically spaced in the range $[10^{-4},1]\unit{cm}$ with 512 points, and the time step equal to $\rm{d}t=0.25\unit{sec}$. 
\begin{figure}[ht]
    \centering
    \includegraphics[width=1\textwidth]{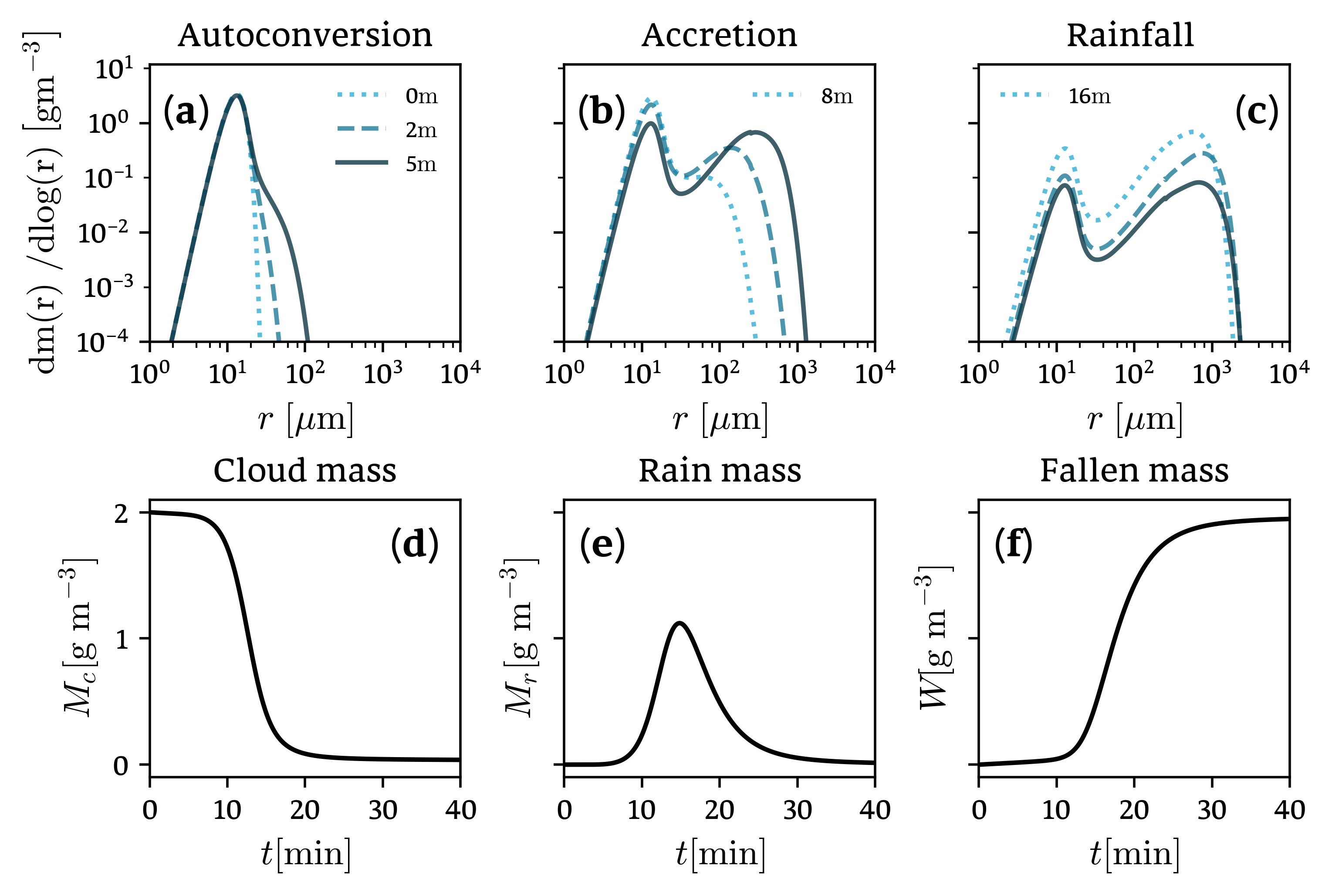}
    \caption[]{\textbf{Top row}: three stages of the bin-resolved cloud dynamics. From left to right these are: I. autoconversion, II. in-cloud mass conserving accretion, and III. simultaneous accretion and rainfall. Each plot shows the mass distribution $\dd m(r)/\dd \log(r)$ as a function of radius, and the three curves in each plot correspond to three different times in the evolution where lighter blues are earlier times. 
    \textbf{Bottom row}: time development of the three bulk variables. From left to right these are cloud mass \Mc, rain mass \Mr, and fallen mass $W$.  Results from a simulation with $L=1\unit{km},\ M_{0}=2\unit{ g/m^{3}},\ N_{0}=400\unit{cm^{-3}},\ r_{0}=12.5\unit{\mu m}$ (no. 24 in Table \ref{table: simulation parameters}) .}
    \label{fig:stages}
\end{figure}

We begin by exploring a representative simulation with $L=1\ \unit{km},\ M_{0}=2\ \unit{ g/m^{3}},\ N_{0}=400\ \unit{cm^{-3}},\ r_{0}=12.5\ \unit{\mu m}$. 
The overall behavior of this simulation is illustrated in Fig. \ref{fig:stages} and follows well-established paradigms~\cite{yau1996short} (see also the animations included in the SI) . Initially, a small population of larger drops slowly forms from collision of the initial drops, a process known as autoconversion (panel a). Later, these larger drops grow more rapidly and to larger radii by catching (accreting)  smaller drops as they fall (panel b). The larger drops eventually grow large enough that they rain out of the cloud, leaving behind a small residual population of unaccreted initial  drops (panel c). The separate peaks in the droplet distribution and their differing behavior due to disparate fall speeds motivates the usual definition of  \emph{cloud drops} as those with $r<40\ \micron$ whose fall speed is neglected, and \emph{rain drops} as those with $r\geq40\ \micron$ whose fall speed is considered. The corresponding mass per unit volume of cloud drops \Mc\ and  rain drops \Mr\  are then straightforwardly defined as
\begin{linenomath}
\begin{subequations}
    \begin{align}
       \Mc(t) & \equiv  \intop_{r<40 \micron} \rho_{w} \frac{4}{3}\pi r^{3}n(r,t)\dd r   \label{eq:Mc_def}  \\
       \Mr(t) & \equiv \intop_{r>40 \micron} \rho_{w} \frac{4}{3}\pi r^{3}n(R,t)\dd R \ . \label{eq:Mr_def}
    \end{align}
    \label{eqs:Mc_Mr_def}
\end{subequations}
\end{linenomath}
 Note that we use $r$ for cloud drop radius and $R$ for rain drop radius, even as dummy variables.
  We also define a third  variable, $W$ the water mass that left the cloud, by the condition that the sum of all three variables is conserved, i.e. 
  \begin{linenomath}
  \begin{align}
  \label{eq:W_def}
        W(t) \equiv M_{0}-M_{r}(t)-M_{c}(t) \ .
  \end{align}
  \end{linenomath}
  The bottom panel of figure \ref{fig:stages} shows the evolution of \Mc, \Mr\ and \W. The same three dynamical stages discussed above can be identified in the bulk dynamics. The very initial stage of autoconversion ($t\sim[0,5]\unit{m}$) is characterized by very small changes in both rain and cloud masses and the absence of fallen mass. Then, during the accretion stage ($t\sim[5,13]\unit{m}$) cloud mass can be seen to decrease, while rain mass increases. All the while, there is still very little fallen mass, \W remaining small, so the in-cloud dynamics are approximately mass conserving. Finally, rainfall onsets as rain mass grows more slowly and eventually declines, accompanied by a steady increase in the fallen mass ($t\sim[15,30]\unit{m}$).
 One novelty of our later formulation of bulk dynamics for \Mc,\Mr\ and \W\ (Section \ref{sec_bulk_dynamics}) will be to approximate the second dynamical stage, accretion (figure \ref{fig:stages}$(b)$), as exactly mass conserving, in contrast to the later stage, accretion and rainfall (figure \ref{fig:stages}$(c)$), in which both the rain drop and cloud drop mass decrease with time. 


\section{Cloud observables} \label{sec_obs}

While the bulk dynamics of $\Mc(t)$ and $\Mr(t)$ will be central in what follows, another aim of this work is to better understand the behavior of certain cloud observables which depend on the microphysical dynamics as a whole and are not functions of time. Specifically, we focus on the precipitation efficiency, cloud lifetime, and average rain rate. We define these quantities in this section and show how they vary with  parameters $(L,M_0,N_0,r_0)$.

\subsection{Observable definitions}
The cloud lifetime $T$ must be defined first, as the other observables depend on $T$. Because the other observables of interest are related to rain,  our focus will be on the accretion and fallout stages of droplet evolution (Fig. \ref{fig:stages}(b),(c)) rather than the autoconversion stage (Fig. \ref{fig:stages}(a)). As such, we measure  $T$  starting from the  time $t_0$ that a significant rain drop population is first present, where the threshold is chosen to be 
\begin{subequations} 
\begin{linenomath}
    \begin{equation} 
        \Delta_i \ \equiv \ \Mr(t_0)/\Mo(t_0) = 0.1
        \label{eq:deltai_def}
    \end{equation}
    \end{linenomath}
to ensure that the dynamics is past the autoconversion stage. Similarly, the end of the lifetime $t_0+T$ is taken to be when the rain drop population has declined and is close to zero, and thus the remaining cloud mass asymptotes to its own final value, as in Fig. \ref{fig:stages}d. The final rain mass threshold is taken to be 
\begin{linenomath}
    \begin{equation}
        \Delta_f \ \equiv \ \Mr(t_0+T)/\Mo(t_0+T) = 0.02,
        \label{eq:deltaf_def}
    \end{equation}
    \label{eq:delta_def}
\end{linenomath}
\end{subequations}
which is sufficiently small to ensure little further evolution of \Mc. 

To define the precipitation efficiency and average rain rate we will need the instantaneous sedimentation flux or rain rate $RR(t)$, obtained as just the mass-weighted integral of the sedimentation term in \eqnref{eq : Smoluchowski in volume}, multiplied by $L$ and with units of mass per unit area per unit time:
\begin{linenomath}
\begin{equation}  
    RR(t) \equiv -\intop \rho_{w} \frac{4}{3}\pi R^{3}u_{g}(R)n(R,t)\dd R \ . \label{eq:RR_def}
\end{equation} 
\end{linenomath}
The total accumulated rain $I$ is then just the time-integral of $RR(t)$ over the cloud lifetime:
\begin{linenomath}
\begin{equation}  
    I \ \equiv \int^{t_0+T}_{t_0} RR(t)\, \dd t =W(t_{0}+T)L\equiv W^{f}L\ .
    \label{eq:I_def}
\end{equation} 
\end{linenomath}
  Note that the the accumulated rain, $I$, is equal to the final value of the fallen mass, $W^{f}$, multiplied by the length of the cloud $L$.
  Sedimentation is the only process by which the total mass can change in our model, so the rain rate we define here can be pictured as the rain rate at cloud base, excluding any rain drop evaporation which might occur in a cloud shaft.

 With the definition \eqref{eq:I_def} of accumulated rain in hand, we can define the precipitation efficiency as simply $I/\LWP$ where $\LWP\equiv \Mo L$ is the initial liquid water path in mass per unit area. The final observable of interest is then the time-averaged rain rate 
 
\begin{equation}
    \RRav \ \equiv \ \frac{1}T\int^{t_0+T}_{t_0} RR(t)\, \dd t =  \frac{I}{T} \ .
\end{equation}

\subsection{Observable and simulation phase space}
With our observables defined, we now examine how they vary as  we sweep parameter space by running 48 simulations with parameters in the range $M_{0}\in [0.75,2] \unit{ g\cdot m^{-3}}$, $N_{0}\in [50,400] \unit{cm^{-3}}$, $r_{0}\in [10,17]\unit{\mu m}$ and $L\in[1,3] \unit{km}$  \cite<similar to the ranges explored in>{seifert_double-moment_2001}. Note that the first three parameters are not independent, therefore not all combinations are possible. The parameter details for all our simulations are presented in the Appendix, Table~\ref{table: simulation parameters}.

\begin{figure}[ht]
    \centering
    \includegraphics[width=1\textwidth]{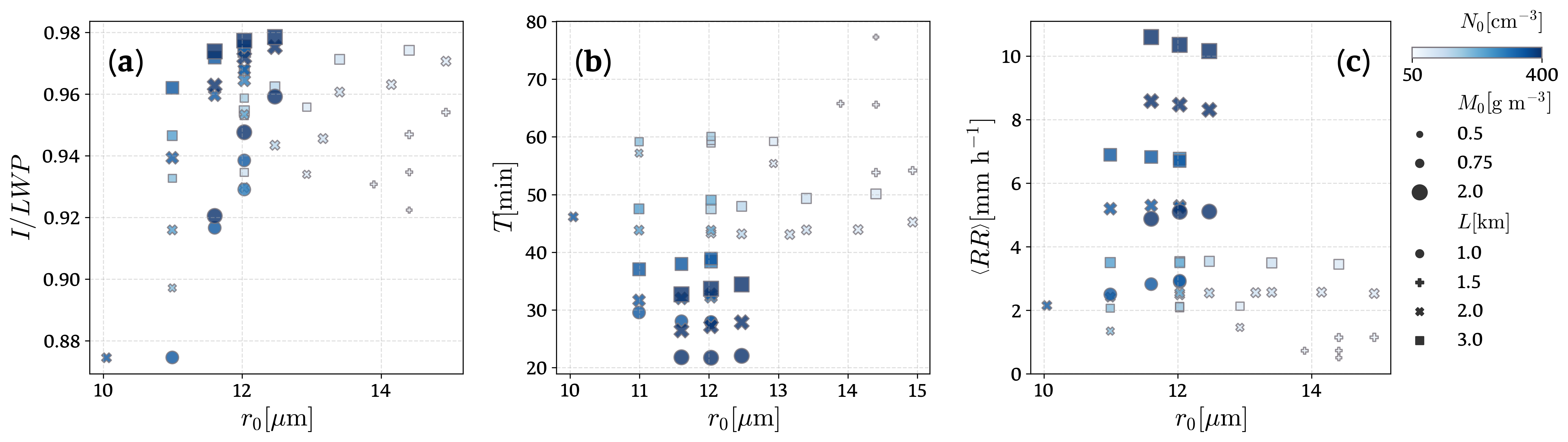}
    \caption{(a) The normalized accumulated rain mass, (b) the average rain rate and (c) lifetime for different clouds vs $r_{0}$. The gradient in color represents the different initial values of $N_{0}$ where lighter blues correspond to a smaller initial number. Different symbols indicate different cloud lengths: $1\unit{km}$ (circle), $1.5\unit{km}$ (cross), $2\unit{km}$ (x) and $3\unit{km}$ (square). The variations in symbol size express the variations in the initial LWC in the range $[0.5,2]\unit{ g \ m^{-3}}$.}
    \label{fig:RR and lifetime}
\end{figure}
We present the precipitation efficiency $I/\LWP$ as a function of $r_0$ in Fig.~\ref{fig:RR and lifetime}(a), where $N_0$ is color-coded, $M_0$ is represented by the size of the markers and the cloud height by different symbols. Figures \ref{fig:RR and lifetime}(b) and (c) similarly show the lifetime $T$ and the rain rate \RRav\ accordingly.

The precipitation efficiency in our simulations ranges between $0.88-0.98$, meaning that for all of our clouds the majority of the mass falls out as rain. However, this result should be taken in the context of our model, which does not include condensation, entrainment, or evaporation of  droplets either within the cloud or below cloud base. Thus, our precipitation efficiency  only accounts for the rain `formation efficiency' \cite{langhans2015a} after some rain drops have already formed, in situations where the collision-coalescence and condensation processes are well separated in time~\cite{dagan_competition_2015}.  In particular, modeling the initial entrainment and mixing during the autoconversion stage is crucial to be able to capture the full range of possible precipitation efficiencies. Indeed, these processes can decrease the initial cloud mass and inhibit the formation of rain drops, thus significantly reducing the precipitation efficiency~\cite{Kessler_on_1969,Seifert_microphysical_2010}.

The lifetimes we find span a relatively wide range, between 20 and 80 minutes. Most of the simulations fall within the physically expected range, though the slowest simulations have slightly longer rain lifetimes than those previously found \cite<e.g.>{jiang2006aerosol}  (such simulations, which have the smallest initial water content \Mo, are probably more characteristic of drizzle in stratocumulus clouds, as can also be seen from their rain rate in panel c). While some rough patterns are evident (e.g. an inverse relation between the lifetime and \Mo), it is difficult to identify any precise relationship between initial parameters and $T$.

The rain rate in our simulations ranges between $0.5-11 \rm mm/h$, in agreement with previous studies for the rain rate at cloud base~\cite{rauber2007rain}. 
Again, some patterns are evident, e.g. a rough inverse relationship between \RRav\ and  $\LWP=\Mo L$, but the scatter is significant and no precise relationships are evident. One of the main results of this paper (section \ref{sec_collapse}) will be to appropriately non-dimensionalize these quantities and identify the non-dimensional parameter which causes this data to collapse.


\section{Bulk dynamics: introduction} \label{sec_bulk_dynamics}
\subsection{Assumptions and formulation}\label{ssec: assum and form}
The next step in understanding our simulations and the scatter evident in Figure \ref{fig:RR and lifetime} is to formulate equations governing the bulk variables \Mc, \Mr and \W (defined in  \eqnref{eqs:Mc_Mr_def} and \eqnref{eq:W_def}).
As mentioned above, we are only interested in the dynamics occurring while a significant rain drop population is present. We thus do not seek to model the non-precipitating, autoconversion stage (Fig.~\ref{fig:stages}a) which produces the nascent rain drop population. (Note that while we do not analytically model the autoconversion phase, we do simulate it; its main impact is to determine the time $t_0$ at which $\Mr=\deltai$.)
 In modeling the remaining droplet dynamics we take a cue from Fig.~\ref{fig:stages} and consider two stages: a \emph{mass-conserving stage}  where  no significant rainfall occurs (Fig.~\ref{fig:stages}b), and a \emph{rainfall stage} in which accretion continues but rain drops are now large enough to fall out of the cloud with a significant rate (Fig.~\ref{fig:stages}c). This is in contrast to parameterization schemes which typically model autoconversion, accretion, and sedimentation simultaneously, using a variety of approaches \cite<e.g.>{seifert_double-moment_2001,reinhardt1974analysis,khairoutdinov2000new,seifert2020potential}.

To model these two stages, we make two assumptions. The first is that accretion of cloud drops by rain drops contribute to \Mcdot\ and $-\Mrdot$ as 
\begin{linenomath}
    \begin{equation} 
        \left.\Mcdot\right|_{\text{ACC}} = - \left.\Mrdot\right|_{\text{ACC}}\ = \ - \krbar \Mr \Mc
        \hspace{1cm} \mbox{(bulk accretion)}
        \label{eq:bulk_acc}
    \end{equation}
\end{linenomath}
where \krbar\ is a constant accretion coefficient. This is a standard and fairly accurate Kessler type parameterization~\cite{Kessler_on_1969}, used e.g. in~\citeA{beheng_parameterization_1994}, which can be derived from \eqref{eq : Smoluchowski in volume} with relatively mild assumptions. We determine  \krbar\ by optimizing the fit between simulated $\Mc(t)$ and that calculated via \eqnref{eq:bulk_acc}. Note that the values so inferred (Fig. \ref{fig : kr value}a) are close to the value $k_r=6 \unit{kg/m^3s}$ appearing in~\cite{beheng_parameterization_1994}. \ref{appendix_acc} provides further discussion and references, a derivation of \eqnref{eq:bulk_acc}, and details on the fit.

The second assumption is that the sedimentation or fallout sink for \Mr\ in the rainfall stage is  approximated, in analogy to the fallout term in Eq. \eqref{eq : Smoluchowski in volume}, as  
\begin{linenomath}
    \begin{equation} 
        \dot{W} = -\left.\Mrdot\right|_{\text{FALL}} \ = \ \frac{\ugbar}{L}\Mr  \ .
        \hspace{1cm} \mbox{(bulk sedimentation)}
        \label{eq:bulk_sed}
    \end{equation}
\end{linenomath}
Here \ugbar\ is a constant (in time) bulk fall speed, again determined separately for each simulation by optimization, but this time we optimize the fit between simulated $W(t)$ and $W(t)$ as calculated via \eqnref{eq:bulk_sed}. This less common parameterization and its validity are  discussed further in Section \ref{sec_sed}, and the fitting procedure is discussed in \ref{appendix_sed}.

With these assumptions, we have the following sets of bulk dynamic equations for the two  stages: 
\begin{subequations} 
\begin{itemize}
    \item Mass-conserving stage:
        \begin{align}
            \dot{M}_{r}=\bar{k}_{r}M_{r}M_{c}, &&
            \dot{M}_{c}=-\bar{k}_{r}M_{r}M_{c}, &&
            \dot{W}=0 \  .
        \label{eq:mass_conserving}
        \end{align}
    

    \item Rainfall stage:
        \begin{align}
            \dot{M}_{r}=\bar{k}_{r}M_{r}M_{c}-\frac{\bar{u}_g}LM_r, &&
            \dot{M}_{c}=-\bar{k}_{r}M_{r}M_{c}, &&
            \dot{W}=\frac{u_{g}}{L}M_{r} \ .
        \label{eq:rainfall}
        \end{align}
\end{itemize}
\label{eq:bulk_dyn}
\end{subequations}
Note that the non-linearity in the resulting bulk parameterization is of the predator-prey type, the rain drops preying on the cloud drops in this analogy. However, in the absence of a source for the cloud drops, the dynamics will not exhibit oscillations between predator and prey. Instead, the rainfall dynamics can be exactly mapped to the so-called SIR model for epidemics ~\cite{kermack1927contribution,kroger2020analytical}.  

\subsection{Non-dimensionalization}
Before solving the bulk dynamic equations \eqref{eq:bulk_dyn} and validating them against our simulations, it is helpful to first non-dimensionalize them. The natural mass scale  is the initial LWC $M_0$, while the natural timescale for the mass-conserving dynamics \eqref{eq:mass_conserving} is the accretion timescale $1/\tauacc\equiv\krbar M_0$.

We accordingly define non-dimensional time and mass fraction variables as 
\begin{linenomath}
\begin{equation}
    \tau \equiv (t-t_0)/\tauacc, \quad \mr \equiv M_r/M_0, \quad  \mc \equiv M_c/M_0, \quad \mbox{and} \quad  w\equiv W/M_{0}.
\end{equation}
\end{linenomath}
Note that $\tau=0$ at $t=t_0$, so $\mr(0)=\Delta_i$ by \eqnref{eq:deltai_def}. In terms of these variables,  the bulk dynamics  \eqref{eq:bulk_dyn} read
\begin{subequations}
    \begin{itemize} 
        \item Mass-conserving stage (non-dimensionalized):
            \begin{align}
                \partial_\tau m_{r}=m_{r}m_{c}, && \partial_\tau m_{c}=-m_{r}m_{c}, && w=0 \ . 
                \label{eq:mass_conserving_re}
            \end{align} 
        \item Rainfall stage (non-dimensionalized):
            \begin{align}
                \partial_\tau m_{r}=(m_c-\mu)m_r, && \partial_\tau m_{c}=-m_{r}m_{c} , && \partial_{\tau}w= \mu m_{r} \ .
                \label{eq:rainfall_re}
            \end{align}
    \end{itemize}
            \label{eq:bulk_dyn_re}
\end{subequations}
Note that the dimensionful rainfall dynamics \eqref{eq:rainfall} contains an additional time-scale, the sedimentation timescale  $1/\taused=\bar{u}_g/L$. The non-dimensionalized rainfall dynamics \eqref{eq:rainfall_re} thus contains a non-dimensional parameter
\begin{linenomath}
\begin{equation}
    \mu \ \equiv \ \tauacc/\taused \ = \frac{\ugbar}{\krbar \Mo L} \ .
    \label{eq:mu_def}
\end{equation}
\end{linenomath}
This parameter will be central in what follows. Also note that the normalized mass conserving dynamics \eqref{eq:mass_conserving_re} contain no parameters. This means  that in the mass-conserving stage, bulk variables exhibit universal behavior once time is measured in accretion time-units \tauacc\ and masses are normalized by the LWC \cite<see also>{srivastava1988}. 
Note that while $1/\tauacc$ sets a typical scale for the accretion rate, from the first equation in    \eqref{eq:mass_conserving_re} $m_c(\tau)$ can be thought of as the instantaneous accretion rate for raindrop growth (in non-dimensionalized units). 
Also, assuming that the dynamics mostly follow the two above stages, with a rapid transition between the two, we see that the peak in the rain mass would be achieved when $m_c\approx \mu$, implying that $\mu$ should be bounded by $1$.

\subsection{$\mu$-dependence}
By the definition \eqref{eq:mu_def} $\mu$ is the ratio of the accretion timescale \tauacc\ to the sedimentation timescale \taused. For smaller $\mu$,  more accretion can happen over a given sedimentation timescale and thus rain drops have more time to capture cloud drops and increase \mr\ before they fall-out. 
More subtly, $\mu$ also sets a threshold for the cloud mass fraction: only when $m_c< \mu$ does sedimentation dominate over accretion for the rain mass and $\partial_\tau m_r<0$. Thus, a cloud mass fraction of at least $1-\mu$ is converted to rain before a depletion of the rain mass can take place.  One thus expects an overall increase in \mr\ and decrease in \mc\ with a decrease in $\mu$, both of the above effects contributing to this trend. Finally, the fact that both the mass-conserving and rainfall regimes only exhibit one free parameter between them suggests that there is a one-parameter family of dynamics, parameterized by $\mu$, and that simulations with different initial conditions but the same $\mu$ should be identical (to the degree that our bulk approximations hold). 

\begin{figure}[t!]
    \centering
    \includegraphics[width=1\textwidth]{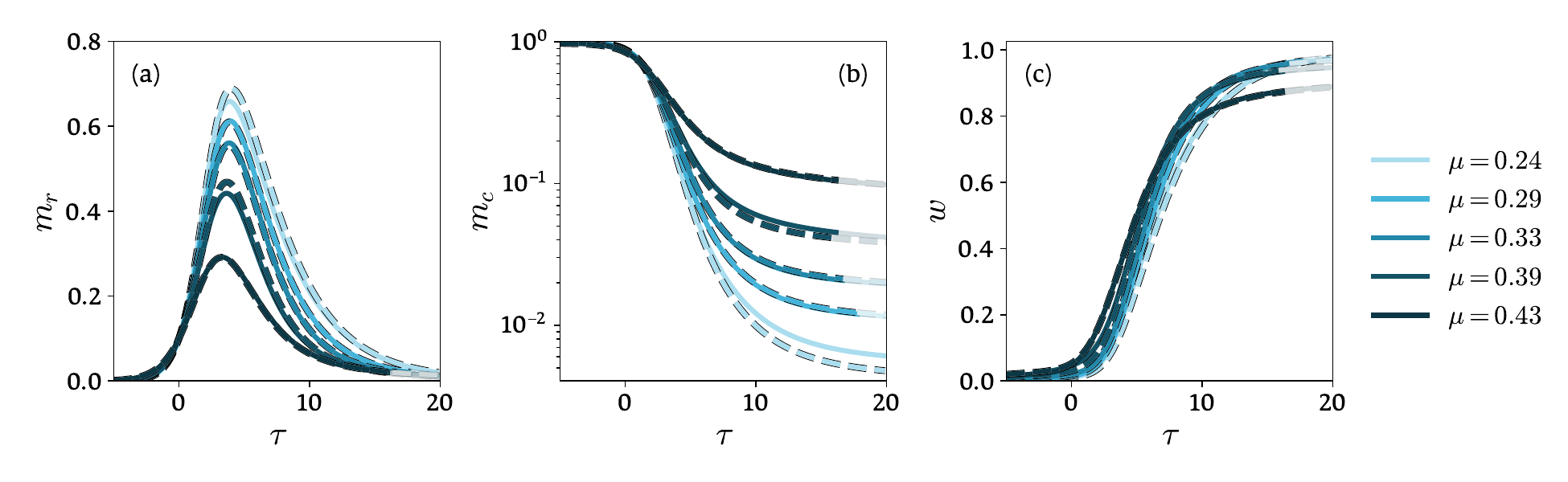}
    \caption{\textit{Normalized in-cloud rain (a) cloud (b) and fallen rain (c) masses, $m_{r},m_{c},w$, as a function of the normalized time $\tau$ shown for ten example cloud simulations as specified in Table \ref{table: simulation parameters of similar mus plot}. Lighter blues indicate smaller $\mu$. The dashed and solid lines represent different simulations with similar $\mu$ values.} Similar $\mu$ values show similar dynamics for all three bulk variables.}
    \label{fig:norm masses}
\end{figure}

To verify these predictions,  Fig. \ref{fig:norm masses}  shows simulated $m_r(\tau)$, $m_c(\tau)$ and $w(\tau)$ for several different values of $\mu$ in different colors, where $\tau$ and $\mu$ are calculated using the inferred \krbar\ and \ugbar\ for each simulation (a typical value for $\mu$ is $\sim 1/3$, a value we discuss further in Section \ref{sec_mu_value}).  For each value of $\mu$ we also present results from two different simulations (solid vs dashed lines), as tabulated in Table~\ref{table: simulation parameters of similar mus plot}. The figure clearly shows that decreases  in $\mu$ yield an increase in \mr\ and a decrease in \mc,   while different simulations with the same $\mu$ essentially collapse on top of each other, as argued above. This is typical  for our simulations, though some outliers do exist; see discussion in \ref{app: comparison of the dynamics}.

\begin{table}[!ht]
\centering
\begin{tabular}{c  c  c  c c c c}
\toprule
        $\mu$&$\LWP[\unit{g m^{-2}}]$& $M_0[\unit{ g\ m^{-3}}]$ & $L[\unit{km}]$&$r_0[\unit{\mu m}]$ & $N_0[\unit{cm^{-3}}]$ & No. \\ \midrule
        0.437 &1.5& 1.5&1&11.0&350&48\\
        0.431 &2 & 1 & 2 & 10.0 & 350&47 \\ \midrule
        0.382 & 4.5 & 1.5 & 3 & 12.0 & 350&38\\ 
        0.387& 0.9 & 0.6 & 1.5 & 14.4 & 60&39 \\ \midrule
        0.327&2.25 & 0.75 & 3 & 12.0 & 175&22 \\ 
        0.331&2 & 2 & 1 & 12.5 & 400&24 \\ \midrule
        0.287&3&1&3&12.0&230&14\\
        0.286 &3&1.5&2&12.0&350&13\\ \midrule
        0.239& 4&2&2&12.5&400&6\\
        0.239&4.5&1.5&3&12.0&300&5\\
  \bottomrule
  \hline
  \end{tabular}
     \caption{Bulk parameters for the simulations presented in Fig.~\ref{fig:norm masses}.}
 \label{table: simulation parameters of similar mus plot} 
\end{table}


\section{Bulk dynamics: solution and validation} \label{sec_bulk_solution}
With the normalized bulk dynamics formulated and general properties explored, we now solve the normalized bulk dynamics \eqref{eq:bulk_dyn_re}.

\subsection{Mass conserving regime}
In the mass conserving stage,  $m_{r}+m_{c}=1$  so  \eqnref{eq:mass_conserving_re} can be reduced to a single ODE and directly solved 
to give: 
\begin{linenomath}
\begin{subequations}
    \begin{eqnarray}
        m_{c}(\tau)&=&\mc(0)\frac{e^{-\tau}}{\mr(0)+\mc(0)e^{-\tau}}
        \label{subeq : mass conserving solutions Mc}\\
        m_{r}(\tau)&=&1-m_{c}(\tau)
        \label{subeq : mass conserving solutions Mr}
    \end{eqnarray}
    \label{eq : mass conserving solutions}
\end{subequations}
\end{linenomath}
where $m_r(0)=\Delta_i$ and $m_c(0)=1-\Delta_i$ are the initial rain mass fraction and cloud mass. This solution should describe the dynamics well from $\tau=0$  until the time when rain-fall begins, i.e. fallout becomes significant. In particular, the dynamics should follow a universal curve, independent of $\mu$ up to that time. 

\begin{figure}
    \includegraphics[width=1\textwidth]{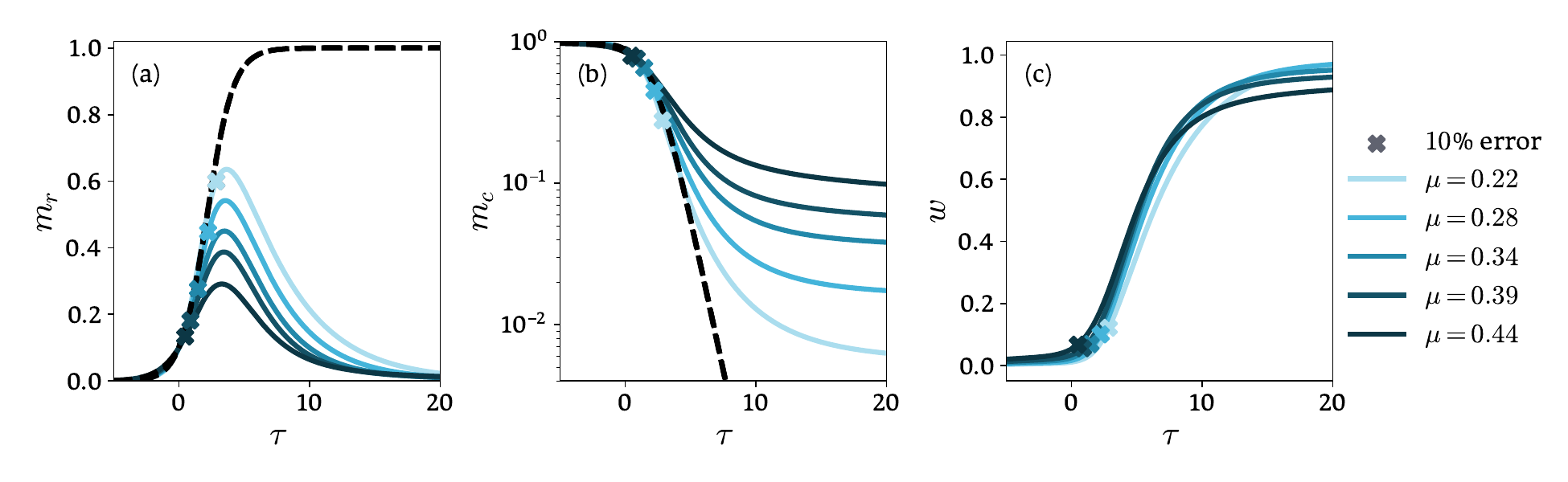}
    \caption{\textit{Normalized in-cloud rain, cloud and accumulated rain masses, $m_{r}$ (a) $,m_{c}$ (b), $w$ (c) as a function of the normalized time $\tau=\bar{k}_{r}M_{0}$ taken from five example cloud simulations, simulations no. 3,11,27,42 and 48 in Table~\ref{table: simulation parameters} where lighter blues indicate smaller $\mu$. The black dashed line represents the in-cloud mass conserving solution \eqref{eq : mass conserving solutions}. The solution for the fallen rain mass in this regime is always zero.} The mass conserving solution is a universal curve which converts, in the long time limit, all of the cloud mass to rain mass as seen in the steady state solution $m_{r}(\tau)=1$.}
    \label{fig: dynamics}
\end{figure}

In Fig.~\ref{fig: dynamics}, we show $w(\tau)$, $m_r(\tau)$, and  $m_c(\tau)$  for several values of $\mu$, and plot in a black dashed line the mass conserving solutions \eqref{eq : mass conserving solutions}. The simulations indeed all follow the dashed curve at early times,  with smaller $\mu$ simulations leaving the dashed curve at later times as sedimentation remains negligible for longer times (relatively larger \taused). The in-cloud mass conserving stage is thus indeed prolonged with decreasing $\mu$, allowing $m_r$ to reach larger values as noted above.

\subsection{Rainfall regime}
We now solve the normalized rainfall stage equations \eqref{eq:rainfall_re} and compare to our simulations. This requires choosing an initial time $\taustar$ to begin their evolution (choosing such a time is also required for the inference of $\bar{u}_g$). While it would be desirable to choose \taustar\ to be before the peak in \mr, such that deviations from the  mass-conserving solutions are still small (i.e before the crosses in Fig. \ref{fig: dynamics}) and the two regimes could be directly connected, this turns out to be too much to hope for: if one does this, errors in the rainfall stage solutions at late times are unacceptably large. As a compromise, we choose \taustar\ to be  \emph{after} the peak in \mr, at the time defined by
\begin{linenomath}
\begin{equation}
    \mr(\taustar) = 0.9\mr^{\text{max}} \ .
    \label{eq:taustar_def}
\end{equation}
\end{linenomath}
We will see below that for larger values of $\mu$ the rainfall regime actually extends to times earlier than \taustar, allowing for some overlap with the mass-conserving regime.

To solve Eqs. \eqref{eq:rainfall_re}, we first rewrite them as
\begin{linenomath}
\begin{align}\label{eq:system}
        \partial_\tau m_{r}=(m_c-\mu)m_r, &&
        \partial_\tau \ln m_{c}=-m_{r} , && \partial_{\tau}w= \mu m_{r} \ .
\end{align}
\end{linenomath}
Remarkably, one can eliminate \mr\ from the last two equations and integrate from \taustar\ forwards to  obtain 
\begin{linenomath}
\begin{equation}
  m_c=m_c(\tau^*)e^{-(w-w(\tau^*))/\mu}  \ .
  \label{eq:mc_formula}
\end{equation}
\end{linenomath}
\begin{figure}[!ht]
    \centering
    \includegraphics[width=0.7\textwidth]{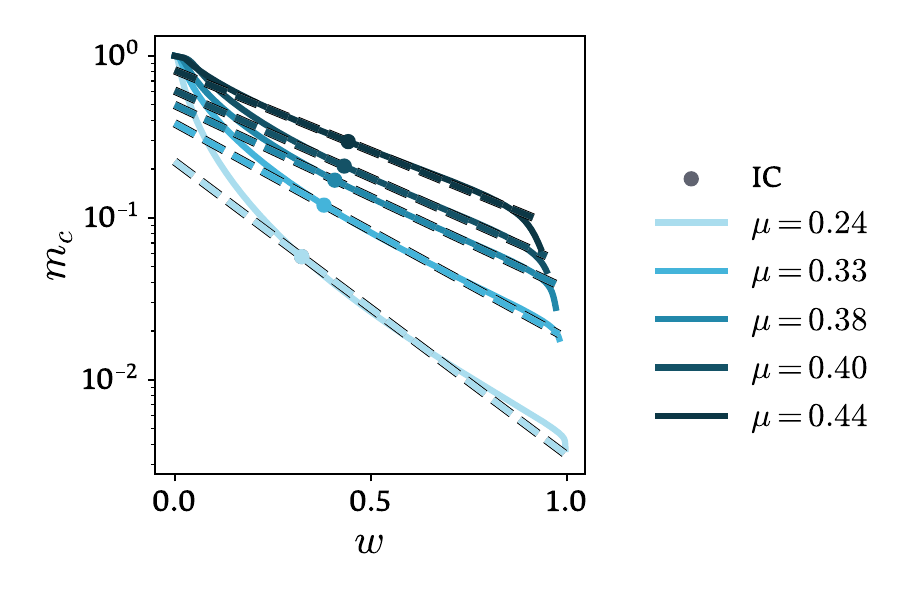}
    \caption{\textit{Normalized cloud mass $m_{c}$ as a function of removed rain mass $w$ taken from five example cloud simulations, simulations no. 5,13,25,39 and 47 in Table~\ref{table: simulation parameters}
    where lighter blues indicate smaller $\mu$. The dashed lines represent the theoretical relation between these two quantities according to \eqref{eq:mc_formula}. The initial condition is taken at $\tau = \tau^{*}$, the time at which we start the fit to infer $\bar{u}_{g}$ and is indicated by a circle.} We indeed see that there is a non negligible range of $w$ at which \eqref{eq:mc_formula} is a good representation for the dynamics. }
    \label{fig: mc_vs_w}
\end{figure}
This direct relation between cloud water \mc\ and accumulated rain $w$ is a novel feature of this analysis. It stems from the fact that \mr\ determines the decrease in \mc\ (via accretion), as well as the increase in $w$ (due to fallout).  Figure \ref{fig: mc_vs_w}  plots simulated $m_c$ as a function of $w$ taken from cloud simulations with five representative values of $\mu$. The time $\tau^*$ is denoted by a circle, and relation \eqref{eq:mc_formula}, with the values of $m_c(\tau^*),w(\tau^*)$ taken from the simulations, is plotted in a dashed line. The excellent agreement between the dashed and solid lines from \taustar\ onwards, except at very late times where $w\sim 1$, validates the relation \eqref{eq:mc_formula}. 

The relation  \eqref{eq:mc_formula} means that we can also express $m_r$ directly in terms of $w$ using the total mass conservation law $w+m_c+m_r=1$:
\begin{linenomath}
\begin{equation}
  m_r(\tau)=1-w-m_c(\tau^*)e^{-(w-w(\tau^*))/\mu}   \ .
  \label{eq:mr_w_star}
\end{equation}
\end{linenomath}
We can plug this expression into the equation for $w$ and solve implicitly for the dynamics of $w$:
\begin{linenomath}
\begin{equation}
  \tau-\tau^{*}=\int_{w(\tau^*)}^{w(\tau)}\frac{\dd w}{\mu\left(1-w-m_c(\tau^*)e^{-(w-w(\tau^*))/\mu}\right)} \ .
  \label{eq:w_integral_star}
\end{equation}
\end{linenomath}

\begin{figure}
    \includegraphics[width=1\textwidth]{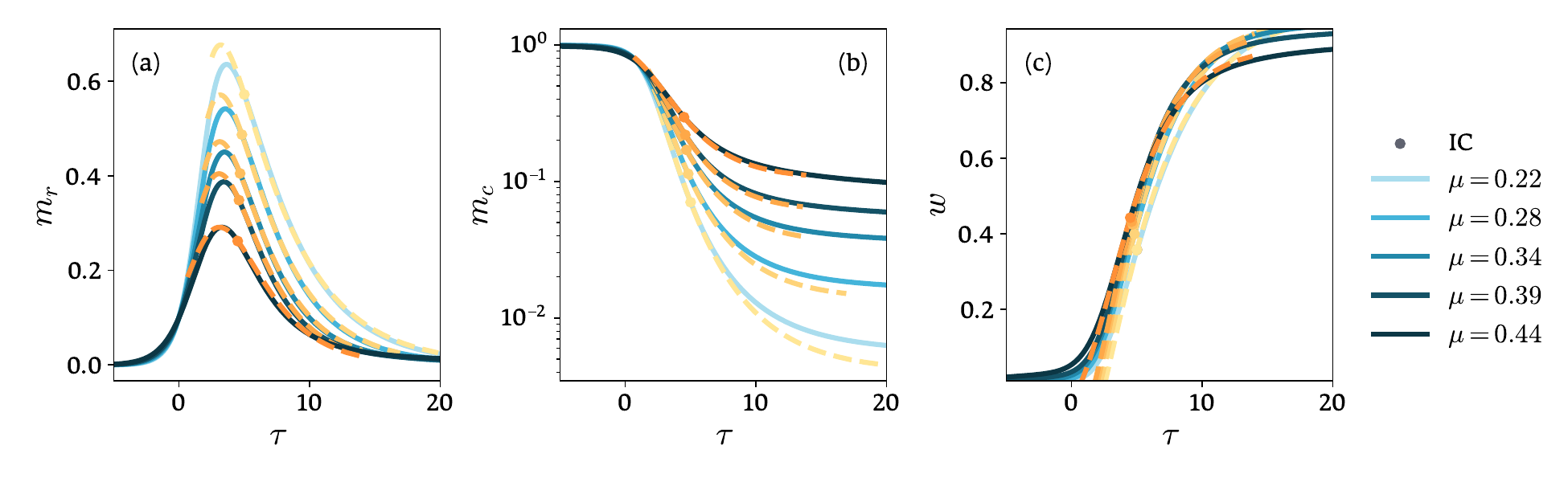}
    \caption{\textit{Normalized in-cloud rain, cloud and accumulated rain masses, $m_{r}$ (a) $,m_{c}$ (b), $w$ (c) as a function of the normalized time $\tau=\bar{k}_{r}M_{0}$ taken from five example cloud simulations, simulations no. 3,11,27,42 and 48 in Table~\ref{table: simulation parameters}
    where lighter blues indicate smaller $\mu$. The orange dashed lines represent the fall-out solutions \eqref{eq:w_integral_star},\eqref{eq:mr_w_star} and \eqref{eq:mc_formula} solved numerically, where the initial condition, indicated by a circle, is taken at $\tau = \tau^{*}$, the time at which we start the fit of $\bar{u}_{g}$}. These solutions are also extended  backwards in time until $w=0$ according to \eqref{eq:w_integral_star}; the accuracy of these extensions  improves with increasing $\mu$.}
    \label{fig: numerics}
\end{figure}

Figure \ref{fig: numerics} plots numerical solutions to \eqref{eq:w_integral_star}, along with the corresponding expressions \eqref{eq:mr_w_star} and \eqref{eq:mc_formula} for \mr\ and \mc, in dashed lines. The initial conditions  $w(\tau^*)$ and $\mc(\taustar))$ are again taken from the simulations. The dashed lines are overlain on simulation results with five representative values of $\mu$ (solid lines). The figure shows rainfall dynamics in the range $w\in[0,w_{f}]$, thus including earlier times $\tau<\taustar$. The rainfall stage expressions \eqref{eq:mc_formula}-\eqref{eq:w_integral_star} do an excellent job of capturing the dynamical evolution in the rainfall stage. Furthermore,  as $\mu$ increases, the rainfall solution is also able to capture the dynamics for $\tau<\taustar$.

This helps explain why our neglect of the transitional stage between the mass-conserving and rainfall stages, when the representative fall-out rate of the rain mass population grows from zero to $\bar{u}_g/L$, does not keep $\mu$ from determining the dynamics (Figure \ref{fig:norm masses}). Figures \ref{fig: dynamics} and \ref{fig: numerics} suggests that this transitional stage is relatively short, as the mass-conserving and rainfall stage together seem to describe most of the time domain, except near the peak in \mr\ for small $\mu$. We will take a closer look at the evolution of the fall-out rate of rain drops in the transitional stage in Section \ref{sec_sed}.

\section{Collapse of observables} \label{sec_collapse}
\begin{figure}
    \includegraphics[width=1\textwidth]{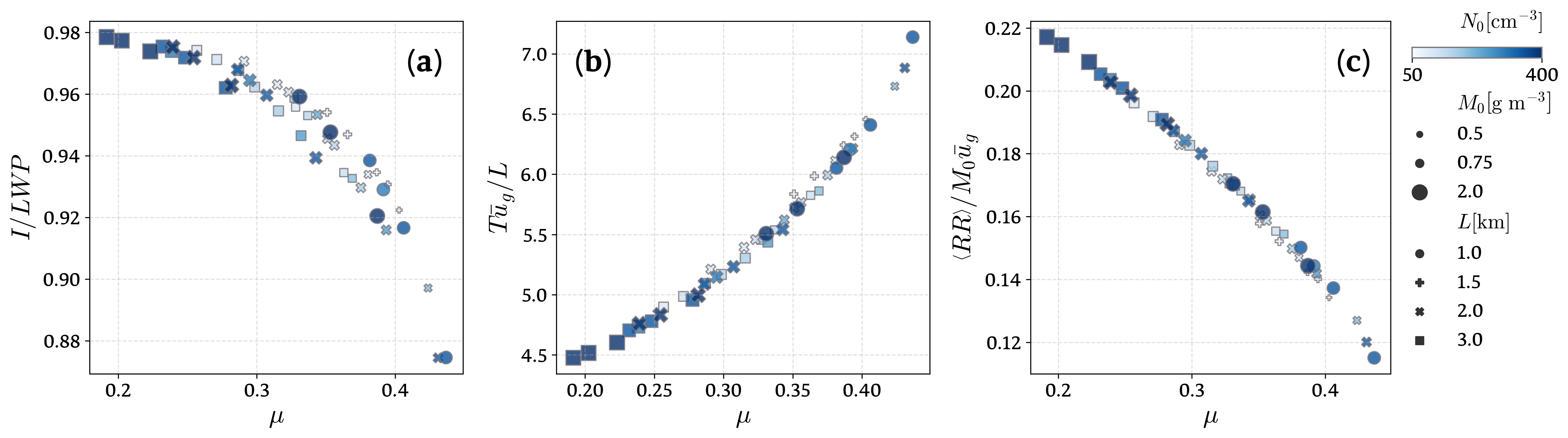}
    \caption{\textit{(a): The accumulated rain mass fraction $I/\LWP=w(\tau_f)=1-\Delta_f-m_c(\tau_f)$ as a function of $\mu$. 
    (b): The lifetime $T\  \bar{u}_{g}/L=\tau\mu$ as a function of $\mu$.
    (c): Average rain rate $\left<RR\right>/M_{0}\bar{u}_{g}$ as a function of $\mu$ for the cloud simulations.
    } All three normalized observables collapse as a function of $\mu$. The range of the precipitation efficiency is relatively small while the normalized lifetime and rain rate span larger ranges.}
    \label{fig: accumulated rain mass normalized}
\end{figure}

Having validated the bulk dynamical description of our simulations, we now return to the cloud observables from Section \ref{sec_obs}. With the parameter $\mu$ in hand and some indication  that the dynamics seem to depend on $\mu$ alone (Fig. \ref{fig:norm masses}), we again plot our observables as in Fig. \ref{fig:RR and lifetime}, but now we normalize the observables and plot them against $\mu$ rather than $r_0$. The result is shown in Figure \ref{fig: accumulated rain mass normalized}. This figure shows that the normalized cloud observables do exhibit a striking  collapse as a function of $\mu$ for all three observables: normalized lifetime, accumulated rain mass and rain rate.
Notice the much smaller range of values for the accumulated rain as compared to those for the lifetime and rain rate. Figure \ref{fig: accumulated rain mass normalized} confirms that our simulations indeed exhibit a one-parameter family of dynamics, governed by $\mu$.

But, what physics determines the overall values of these normalized observables, and their muted but non-negligible dependence on $\mu$? We address these questions separately for each cloud observable, relying on the analytical solutions from the previous section. Note that our three observables are not independent since the rain rate \RRav\ is given by the ratio between the accumulated mass fraction $I$ and the cloud lifetime $T$. Thus, it is sufficient to focus on the two latter. 

Our calculations in this section will be back-of-the-envelope; a more precise analytical calculation of observables using the bulk dynamics, along with quantitative comparison to the simulations, is given in\ref{app: analytical approx}.

\subsection{Accumulated rain mass}
As seen in figure \ref{fig: accumulated rain mass normalized}a, the normalized accumulated rain mass has a characteristic value of roughly $I/\LWP\approx 0.95$, and decreases monotonically with $\mu$ (we commented on the large value of $I/\LWP$ above). To explain these features, we first appeal to the formula \eqref{eq:mc_formula} for $\mc(w)$ to estimate $\mc(\tauf)$. For the sake of this estimate, we here choose \taustar=\taupeak\, where \taupeak\ is the time at which $\mc=\mu$ and hence \mr\ maximizes (\eqnref{eq:bulk_dyn_re}). We also assume $w(\taupeak)\approx 0$. We are thus implicitly assuming that the rainfall dynamics describe the behavior at the peak in \mr and that the in-cloud mass is conserved prior to the peak.
We further assume that at the final time \tauf\ we may take  $w(\tauf) \approx 1$ in \eqref{eq:mc_formula} ($w(\tauf)$  is ultimately what we seek to estimate, but its difference from 1 is unimportant for determining $\mc(\tauf)$). These are all crude assumptions, but are expedient for the calculation below.

With these assumptions, and assuming a characteristic value $\mu\approx 1/3$ (Table \ref{table: simulation parameters}), \eqnref{eq:mc_formula} becomes simply
\begin{linenomath}
\begin{equation}
    \mc(\tauf) \ = \  \mu e^{-1/\mu}\  \approx \ 0.017\ .
    \label{eq:mcf}
\end{equation}
\end{linenomath}
This equation embodies two key insights: first, that in order for the cloud to enter the rainfall regime the mass of the cloud drops must decrease to a value of order $\mu$. Second, that in the rainfall regime \mc\ declines exponentially by accretion with timescale \tauacc, but only  while there is a significant rain drop population. Accretion is thus only effective over a  nondimensional timescale \taused, before significant rain has fallen out. Once \taused\ has elapsed in the rainfall regime, the remaining cloud droplet mass is $m_{c}(\tau^{*}) e^{-\taused/\tauacc}=\mu e^{-1/\mu}$ and this mass remains in the cloud essentially forever, because the predator rain drops have fallen out. 

With $\mc(\tauf)$ in hand we straightforwarldy obtain \wf\ by  evaluating the definition $w=1-\mc-\mr$ at \tauf, recalling that $m_r(\tau_f)=\Delta_f$:
\begin{linenomath}
\begin{equation}
    w(\tau_f) =1-m_r(\tau_f)-m_c(\tau_f)=1-\Delta_f- \mu e^{-1/\mu} \ \approx 0.96  \ ,
    \label{eq:wf}
\end{equation}
\end{linenomath}
consistent with the values of $I/\LWP$ in Fig. \ref{fig: accumulated rain mass normalized}($a$). Equation \eqref{eq:wf} also qualitatively captures the slight  variations with $\mu$ seen in this figure ($\pm 5\%$ or so): due to the $\mu$-dependence of $\mc(\tauf)$, as $\mu$ increases the cloud mass can before the rainfall stage increases and the amount of accretion occurring over the sedimentation timescale decreases, leaving more unaccreted \mc\ behind and decreasing the accumulated water $I/\LWP\approx w(\tauf)$. Note, however, that equation \eqnref{eq:wf} is too crude an approximation to quantitatively capture the  $\mu$-dependence in Fig. \ref{fig: accumulated rain mass normalized}$(a)$; a better approximation is discussed in \ref{app: analytical approx}.

\subsection{Cloud lifetime}
To understand the dependence of the cloud lifetime on $\mu$ first recall that in accretion time-units, the sedimentation time-scale is given by $1/\mu$. Furthermore, on general grounds we expect $\mu<1$ meaning that the sedimentation regime should be longer than the mass-conserving regime (as can be seen in  Figure \ref{fig: numerics}(b)), so we expect its duration to dominate the cloud lifetime. We thus consider a lifetime normalized by \taused\ rather than \tauacc, i.e. 
\begin{linenomath}
\begin{equation}
    T\frac{\bar{u}_g}{L}=\mu \tau_f \ .
    \label{eq:lifetime_re}
\end{equation}
\end{linenomath}
It is this normalized lifetime which is shown in Figure \ref{fig: accumulated rain mass normalized}b. Note that its variations are indeed small ($\pm 10-15\%$), and much reduced relative to the un-normalized lifetime (Fig. \ref{fig:RR and lifetime}), indicating the appropriateness of normalizing by \taused\ rather than \tauacc. 

The normalized cloud lifetime \eqref{eq:lifetime_re} has a characteristic value of roughly 5, and increases somewhat with $\mu$.
To understand these features we will estimate separately the lifetime of the mass-conserving and rainfall regime as \taupeak\  and \tauf-\taupeak\ respectively. We can  then sum these separate lifetimes and multiply by $\mu$ to evaluate  \eqref{eq:lifetime_re}.
  
We estimate $\taupeak$ from the mass-conserving solution \eqref{subeq : mass conserving solutions Mc} by setting $m_{c}(\tau_{peak})=\mu$, $m_r(\tau_0)=\Delta_i$, and $m_c(\tau_0)=1-\Delta_i$ and solving for $\tau$, yielding
\begin{linenomath}
\begin{equation}
    \tau_{peak} \ = \ \ln\left[\frac{(1-\Delta_{i})(1-\mu)}{\mu \Delta_{i}}\right] \  \approx \ 3\ , \  
    \label{eq:lifetime_mass_conserving}
\end{equation}
\end{linenomath}
which is in the ballpark of \taupeak\ as inferred from Fig. \ref{fig: dynamics}.

To estimate the duration of the sedimentation stage \tauf-\taupeak, ending when $m_r=\Delta_f$, we shall take $m_c$ to be fixed to its final value throughout the sedimentation stage, $\mc=\mc(\tauf)$, in the \mr\ equation in \eqnref{eq:rainfall_re}. Integrating then yields
\begin{linenomath}
\begin{equation}
    \mr(\tau) \approx \mr(\taupeak)e^{-(\mu-\mc(\tauf))(\tau-\taupeak)}
    \label{eq:mr_approx_soln}
\end{equation}
\end{linenomath}

If $\Delta_f$ is sufficiently small, we expect the duration of sedimentation to be dominated by this exponential tail away from the peak of $m_r$, and thus for this to approximation to capture the main dependencies, see SI. 
To compensate for our underestimation due to the use of $m_c(\tau_f)$, we overestimate the rain mass as $m_r(\tau_{\text{peak}})=1$, instead of $m_r(\tau_{\text{peak}})=1-\mu$ as our assumption of in-cloud mass conservation at that time would have required. With this assumption and invoking \eqnref{eq:mcf}, we can evaluate  \eqnref{eq:mr_approx_soln} at $\tauf$ and rearrange to obtain
\begin{linenomath}
\begin{equation}
     \tau_f-\tau_{peak}=\frac{1}{\mu(1-e^{-1/\mu})}\ln\left[\frac{1}{\Delta_{f}}\right] \ \approx \ 12 \ ,
\end{equation}
\end{linenomath}
roughly consistent with Fig. \ref{fig: numerics} and indeed much larger than the mass-conserving lifetime \eqref{eq:lifetime_mass_conserving}.

With the two lifetimes in place, the total normalized lifetime \eqref{eq:lifetime_re} is then
\begin{linenomath}
\begin{equation}
    T\frac{\bar{u}_g}{L}\ \approx \ \mu \ln\left[\frac{(1-\Delta_{i})(1-\mu)}{\mu \Delta_{i}}\right] +\frac{1}{1- e^{-1/\mu}}\ln\left[\frac{1}{\Delta_{f}}\right] \ \approx \ 5,
\end{equation}
\label{eq: normalized lifetime}
\end{linenomath}
roughly consistent with Fig. \ref{fig: accumulated rain mass normalized}(b). The slight $\mu$ dependence of the normalized lifetime stems from the mass conserving regime (giving a linear dependence on $\mu$) as well as a contribution from the rainfall lifetime, related to the $e^{-1/\mu}$ term representing the final cloud mass \eqref{eq:mcf}. Note however, that in \eqnref{eq: normalized lifetime} this term has a negligible effect. A better approximation, which brings this dependence to light, as well as more systematic derivations and detailed comparisons with simulations are shown in the \ref{app: analytical approx}.

\subsection{Rain rate}
If we divide the normalized accumulated rain water $I/\LWP$ by the normalized cloud lifetime $T\ugbar/L$, we obtain the normalized rain rate $\langle RR\rangle /(M_0\bar{u}_g)$, which was plotted in Fig. \ref{fig: accumulated rain mass normalized}c. As a ratio of the previous two cloud observables, it inherits its mean values and variations from them: a characteristic value is $0.97/5=0.2$, and the $\pm 20\%$ variations with $\mu$ stem primarily from similar variations in the normalized lifetime.


\section{Loose ends} \label{sec_loose_ends}

\subsection{Value of $\mu$} \label{sec_mu_value}
We noted above that a typical value of $\mu$ is roughly 1/3. From \eqnref{eq:rainfall_re}, it is also clear that a transition from accretion domination $\mc>\mu$ to rainfall domination $\mc<\mu$ requires $\mu < 1$. But what sets this value?

To estimate $\mu$, we assume that a well-separated raindrop population develops, and we invoke the approximation $\krbar = 0.75 k_3/\rho_{w}$ [cf. \eqnref{eq:krbar_approx}]  as well as the fallspeed parameterization $\ugbar=k_3 \bar{R}$ (cf. \eqref{eq:ug(R)}), with $k_3=8\times 10^3\ \sec^{-1}$. We take $\LWP=L\Mo=2\ \kg/\meter^2$ as a typical value and  assume an average raindrop radius of $\bar{R}=0.6\ \mm$, 
which yields $\ugbar\approx 5 \ \meter/\sec$, consistent with the inferred values in Fig. \ref{fig : ug value} below for this \LWP. With this we have
\begin{linenomath}
\begin{equation}
    \mu \ = \ \frac{\ugbar/L}{\krbar \Mo} \ = \ \frac{4}{3}\frac{\bar{R\rho_{w}}}{L\Mo} \ \approx \ 0.4 \ ,
    \label{eq:mu_estimate}
\end{equation}
\end{linenomath}
a decent ballpark estimate. Note that capturing the variations in $\mu$ across simulations, as required for the collapse of the data in Fig.~\ref{fig: accumulated rain mass normalized}, requires the inference of both $\bar{u}_g$ and $\krbar$ as discussed in
 \ref{subsec : static mu collapse}.

Equation \ref{eq:mu_estimate} suggests that $\LWP=L\Mo$ is a strong control on $\mu$. This is indeed true, as is evident from the strong dependence of $\bar{u}_g$ on the LWP seen in Fig.~\ref{fig : ug value} in \ref{appendix_sed}, but $\bar{R}$ may vary in an uncontrolled way: e.g. decreasing $L$ decreases the fallout rate uniformly for all drops, allowing the transition to rainfall to occur at a smaller $\bar{R}$. Furthermore, several approximations were made in the derivation of \eqnref{eq:mu_estimate} which may break down to varying degrees in different simulations. Alternatively, by the definition \eqref{eq:mu_def},  $\mu$ depends on the inferred parameters  $\bar{u}_g$ and $\bar{k}_r$ whose dependence on the initial parameters $(L,r_0,N_0,M_0)$ is non-trivial and not entirely clear (cf. Figs. \ref{fig : kr value} and \ref{fig : ug value}; note however that \ugbar\ does have a strong dependence on \LWP). From either view, then, it is clear that  $\mu$ should be viewed as an \emph{emergent} parameter, rather than a control parameter.

Furthermore, and as an illustration of this, if we try to increase $\mu$ further, by (say) either decreasing $L$ or $M_0$, the development of large rain drops becomes inhibited. 
This is either because accretion is so slow or because the cloud is so short that whenever an intermediate size drop is created, it immediately falls out of the cloud, not allowing for the growth of a rain drop population. As our scheme does not include evaporation, the cloud population keeps (slowly) developing over very long timescales. So much so, that the sedimentation of \emph{cloud} drops eventually becomes a relevant process. In these extreme cases, a separation into a cloud and rain population no longer makes sense, and the framework developed here no longer applies.

\subsection{Time-dependent bulk fall speed} \label{sec_sed}
The main assumptions in our analysis  (Section \ref{sec_bulk_dynamics}) are that we can model the dynamics in two stages, a mass-conserving stage and a rainfall stage, the latter of which exhibits a constant  (in time) bulk fall speed for the rain water mass. In the simulations, of course, one can diagnose an instantaneous, mass-weighted average fall speed $\ugbar(t)$ as 
\begin{linenomath}
\begin{equation}
    \bar{u}_g(t)\ \equiv \ \frac{\intop_{\text{rain}} \rho_{w}\frac{4}{3}\pi R^{3}n(R,t) u_{g}(R)\dd R}{M_{r}}
    \label{def:avg ug}
\end{equation}
\end{linenomath}
(this is also just the rain rate \eqref{eq:RR_def} divided by \Mr). This definition is equivalent to the effective terminal speed defined in~\cite{koren_aerosol_2015}. This fall speed should smoothly ramp up from near zero to a peak value as the raindrop population develops and moves to larger radii, and should then smoothly decline as the larger drops fallout and the raindrop population declines. How do we reconcile such a time-dependent fall speed with the success of the constant fall speed assumption, which appears to capture the evolution of $\mr(\tau)$ quite well in the rainfall stage (Fig. \ref{fig: numerics}(b)) and also correctly predicts that $\mu$ governs the dynamics?

\begin{figure}[ht]
\noindent\includegraphics[width=0.7\textwidth]{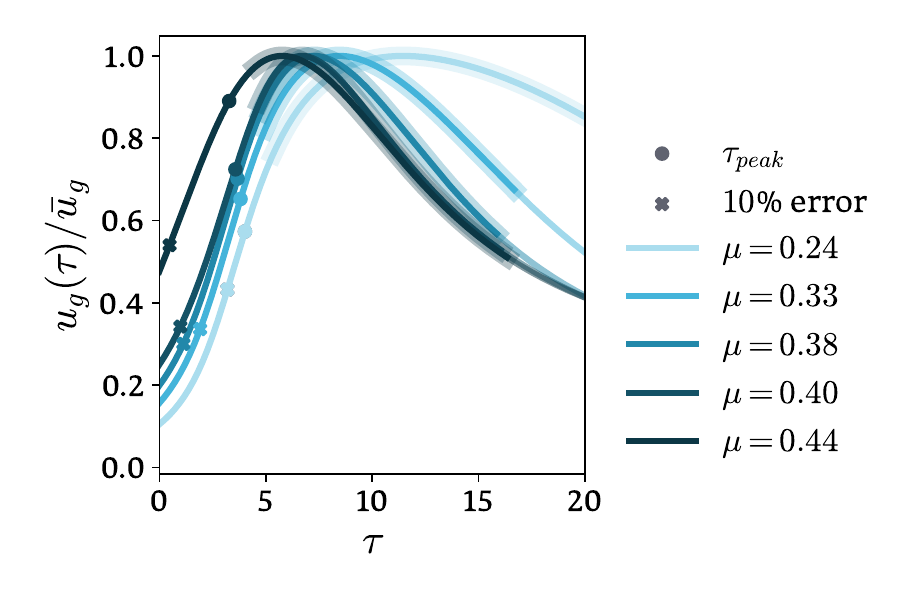}
  \caption{\textit{$\ugbar(t)$ as defined in \eqref{def:avg ug} normalized by the inferred \ugbar\ as a function of the normalized time $\tau=k_{r}M_{0}t$ taken from five simulations, numbers 5,13,25,39 and 47 in Table~\ref{table: simulation parameters}
    where lighter blues indicate smaller $\mu$.} The time at which $\dot{M}_{r}(\tau_{peak})=0$ is denoted on the plot by a dot. This is the time at which the rates switch, and fall-out becomes the dominant process for the rain mass.  We also denote the first time at which the mass conserving solution \eqref{eq:mass_conserving} deviates from the simulation by 10\%, measured according to the rain mass, $M_{r}$. The length of the fit is symbolized with a glow. }
  \label{fig : ug dynamics}
\end{figure}

To explore the behaviour of the sedimentation speed,  Fig. \ref{fig : ug dynamics} shows the temporal evolution of  $\ugbar(\tau)$, normalized by the inferred constant value \ugbar. Results from five  simulations with representative values of $\mu$ are shown. In all cases, $\ugbar(\tau)$ is seen to first increase, then peak at a maximal value \emph{very close to the inferred} \ugbar, and subsequently decrease, as expected. The peak in $\ugbar(\tau)$ occurs slightly after \taupeak. We mark by an `x' the time when the cloud simulations deviate from the mass conserving dynamics by $10\%$ (measured for $m_r$), by a dot the time when $m_r$ reaches its peak, and by a thick translucent line the temporal interval over which $\bar{u}_g$ is inferred (starting at $\tau^*$). 

The reason the bulk rainfall equation \eqref{eq:rainfall_re} can predict $\mr(\tau)$ in the rainfall stage despite a significant decline  in $\ugbar(\tau)$ seems to be that \ugbar\ is a good approximation (i.e. within 20\%\ or so) to $\ugbar(\tau)$ for $5<\tau<10$, during which \mr\ declines exponentially from its peak values by a factor of $\sim \exp(-5\mu) = 0.2$ [cf. \eqnref{eq:mr_approx_soln}]. From Fig. \ref{fig: numerics}b, further declines in \mr\ are rather insignificant, except in the small $\mu$ cases where (fortunately) the decline in $\ugbar(\tau)$ is much slower and the inferred \ugbar\ remains a good approximation for longer. In short, most fallout occurs  near the peak in $\ugbar(\tau)$, which is when fallout is most effective, so values of $\ugbar(\tau)$ far from the peak are not as relevant. 

\section{Conclusion}
This work has shown that:
\begin{itemize}
    \item The dynamics of gravitational coagulation and fallout are governed by a single parameter $\mu$, the ratio of accretion and sedimentation timescales
    \item This ratio determines how long rain drops may prey on cloud drops by accretion, before themselves succumbing to sedimentation
    \item It also sets a threshold on how much cloud mass needs to be converted to rain mass before the latter starts declining
    \item A constant bulk fall speed performs surprisingly well in capturing the bulk dynamics of sedimentation
\end{itemize}

We also found an intriguing  direct relationship between cloud water and accumulated rain [\eqnref{eq:mc_formula} and Fig. \ref{fig: mc_vs_w}]. Future work could further explore the implications of this relationship and potential applications.

As discussed in the earlier sections of the paper, the system we study here is idealized in a number of ways. We do not consider condensation, turbulence, entrainment, or evaporation, allowing us to focus on the two simplest processes in a clouds dynamics: accretion and sedimentation. Our findings can thus be directly relevant only for a part of a clouds life cycle and only for clouds where these processes are well separated in time, with no significant condensation and evaporation occurring during accretion and rainfall. Clouds in a mildly polluted environment seem to be one such example~\cite{dagan_competition_2015}, and are a natural starting point to test and expand upon our work in a more comprehensive modeling context where these additional processes are included. 
More broadly, however, the limitations of our modeling are reflected in the cloud observables:  although we obtain realistic cloud lifetimes and rain rates, the precipitation efficiencies we obtain ($\sim 90\%)$ are very large as compared  to those obtained from other kinds of models, particularly 3D cloud-resolving fluid-dynamical simulations~\cite{Chung_precipitation_2020,lutsko2023}. However, it is worth noting that our precipitation efficiency is just the  `formation' or `conversion' efficiency of cloud drops to rain drops, which is only one factor in the overall precipitation efficiency and which on its own can be as high as 80\% \cite{langhans2015a,lutsko2018b,jeevanjee2022}. 
Nonetheless, future work could investigate whether our overestimate of precipitation efficiency  is indeed due mostly to entrainment at the early stages of the rain population development, and if other effective dimensionless parameters could be introduced to capture it. 

From a broader view, the success of cloud parameterizations requires the existence of an effective coarse-grained model. Indeed, bulk parameterizations are coarse-grained versions of the detailed microphysics dynamics in spectral bin schemes. An even coarser cloud parameterization is a classification of clouds according to their outcomes: mapping bulk cloud initial conditions to time-averaged or cumulative observables. However, the existence of such a reduced description is far from being guaranteed. Here, building on all three levels of coarse graining, we have shown how a minimal effective description can emerge.  While the simplicity of our modeling certainly limits the direct applicability of these results, we suggest our approach as a systematic way to study cloud parameterizations going forward. This raises questions, however, about the appropriate modeling framework for such work. It requires minimal models, detailed enough to capture all the relevant physical processes, but not too detailed to be overly complex or specific. Recent idealized studies take a variety of approaches \cite{Resin_rain_1996,morrison_comparison_2007,Stevens_understanding_2008,Seifert_microphysical_2010,khain_representation_2015},
whose outcomes would have to be compared and contrasted to develop a unified understanding. A broader aim of this study, beyond our specific findings, is therefore to motivate additional idealized work in cloud microphysics, to help build the kind of hierarchical understanding of the subject which the present moment and current applications require.


\clearpage
\appendix
\section{simulation parameters}
\begin{table}[!ht]
    \centering
    \begin{tabular}{l|llllll}
        \toprule
         $\# $&$\LWP[\unit{ g\ m^{-2}}]$&$M_0[\unit{ g\ m^{-3}}]$ & $L[\unit{km}]$ &$r_0[\unit{\mu m}]$  & $N_0[\unit{cm^{-3}}]$&$\mu$ \\ \midrule
        1 & 6 & 2 & 3 & 12.5 & 400 & 0.191 \\ 
        2 & 6 & 2 & 3 & 12.0 & 400 & 0.202 \\ 
        3 & 6 & 2 & 3 & 11.6 & 400 & 0.223 \\ 
        4 & 4.5 & 1.5 & 3 & 12.0 & 350 & 0.232 \\ 
        5 & 4.5 & 1.5 & 3 & 12.0 & 300 & 0.239 \\ 
        6 & 4 & 2 & 2 & 12.5 & 400 & 0.239 \\ 
        7 & 4.5 & 1.5 & 3 & 11.6 & 350 & 0.248 \\ 
        8 & 4 & 2 & 2 & 12.0 & 400 & 0.254 \\ 
        9 & 3 & 1 & 3 & 14.4 & 100 & 0.257 \\ 
        10 & 3 & 1 & 3 & 13.4 & 125 & 0.271 \\ 
        11 & 4.5 & 1.5 & 3 & 11.0 & 350 & 0.278 \\ 
        12 & 4 & 2 & 2 & 11.6 & 400 & 0.282 \\ 
        13 & 3 & 1.5 & 2 & 12.0 & 350 & 0.286 \\ 
        14 & 3 & 1 & 3 & 12.0 & 230 & 0.287 \\ 
        15 & 2 & 1 & 2 & 14.9 & 100 & 0.291 \\ 
        16 & 3 & 1.5 & 2 & 12.0 & 300 & 0.295 \\ 
        17 & 3 & 1 & 3 & 12.5 & 150 & 0.298 \\ 
        18 & 3 & 1.5 & 2 & 11.6 & 350 & 0.307 \\ 
        19 & 2 & 1 & 2 & 14.1 & 100 & 0.315 \\ 
        20 & 3 & 1 & 3 & 12.0 & 170 & 0.316 \\ 
        21 & 2 & 1 & 2 & 13.4 & 125 & 0.323 \\ 
        22 & 2.25 & 0.75 & 3 & 12.0 & 175 & 0.327 \\ 
        23 & 2.25 & 0.75 & 3 & 12.9 & 100 & 0.328 \\ 
        24 & 2 & 2 & 1 & 12.5 & 400 & 0.331 \\ 
        25 & 3 & 1 & 3 & 11.0 & 250 & 0.332 \\ 
        26 & 2.25 & 0.75 & 3 & 12.0 & 150 & 0.337 \\ 
        27 & 3 & 1.5 & 2 & 11.0 & 350 & 0.343 \\ 
        28 & 2 & 1 & 2 & 12.0 & 230 & 0.344 \\ 
        29 & 1.125 & 0.75 & 1.5 & 14.9 & 70 & 0.351 \\ 
        30 & 2 & 1 & 2 & 13.2 & 125 & 0.351 \\ 
        31 & 2 & 2 & 1 & 12.0 & 400 & 0.353 \\ 
        32 & 2 & 1 & 2 & 12.5 & 150 & 0.356 \\ 
        33 & 2.25 & 0.75 & 3 & 12.0 & 125 & 0.363 \\ 
        34 & 1.125 & 0.75 & 1.5 & 14.4 & 75 & 0.366 \\ 
        35 & 2.25 & 0.75 & 3 & 11.0 & 200 & 0.369 \\ 
        36 & 2 & 1 & 2 & 12.0 & 170 & 0.375 \\ 
        37 & 1.5 & 0.75 & 2 & 12.9 & 100 & 0.380 \\ 
        38 & 1.5 & 1.5 & 1 & 12.0 & 350 & 0.382 \\ 
        39 & 0.9 & 0.6 & 1.5 & 14.4 & 60 & 0.387 \\ 
        40 & 2 & 2 & 1 & 11.6 & 400 & 0.387 \\ 
        41 & 1.5 & 1.5 & 1 & 12.0 & 300 & 0.391 \\ 
        42 & 2 & 1 & 2 & 11.0 & 250 & 0.393 \\ 
        43 & 0.9 & 0.6 & 1.5 & 13.9 & 70 & 0.395 \\ 
        44 & 0.75 & 0.5 & 1.5 & 14.4 & 50 & 0.403 \\ 
        45 & 1.5 & 1.5 & 1 & 11.6 & 350 & 0.406 \\ 
        46 & 1.5 & 0.75 & 2 & 11.0 & 200 & 0.424 \\ 
        47 & 2 & 1 & 2 & 10.0 & 350 & 0.431 \\ 
        48 & 1.5 & 1.5 & 1 & 11.0 & 350 & 0.437 \\ 
               \bottomrule
    \end{tabular}
      \caption{Simulation parameters}
  \label{table: simulation parameters} 
\end{table}


\section{Accretion: parameterization and inference}\label{appendix_acc}
\subsection*{Derivation of bulk accretion parameterization}
The form of the accretion parameterization \eqref{eq:bulk_acc} can be derived straightforwardly from the Smoluchowski equation \eqref{eq : Smoluchowski in volume}. Since we are considering cloud drops of radius $r<40\ \micron$, we drop the first term on the right-hand side of \eqref{eq : Smoluchowski in volume} as sedimentation is negligible, and we drop the second term as we neglect autoconversion. Keeping only the third term and integrating over cloud mass (i.e. over $0<r<40\ \micron)$ then yields
\begin{linenomath}
\begin{equation}
    \begin{aligned}
    \dot{M}_c= -\intop_0^{40\rm \mu m} \rho_{w}\frac{4}{3}\pi r^{3} n(r)\dd r\intop_{40\mu m}^{\infty} H(R,r)n(R)\dd R \ .
    \end{aligned}
    \label{eq:accr}
\end{equation}
\end{linenomath}
We next simplify the coagulation kernel $H(R,r)$ by the standard argument that cloud drops are much smaller than rain drops, $r\ll R$, have a collision efficiency $E(R,r)\approx 1$, and again  have negligible fallspeeds $u_g(r)\approx 0$. With these approximations the coagulation kernel \eqref{def : gravitational kernel} simplifies to
\begin{linenomath}
\begin{equation}
    H(R,r) \approx \pi R^{2}\left|u_{g}(R)\right|
    \label{eq:kernel_approx1}
\end{equation}
\end{linenomath}
which is just the volume swept out per unit time per rain drop. Now recall the approximate form for the terminal velocity for rain drops \cite{yau1996short}: 
\begin{linenomath}
\begin{align} u_g(R)\approx
    \begin{cases}
        k_{3}R  \qquad 40\unit{\mu m}<R<0.6\unit{mm} \\
        k_2 \sqrt{R} \qquad R> 0.6 \rm mm
    \end{cases}
\label{eq:ug(R)}
\end{align}
\end{linenomath}
where $k_3=8\times 10^3 \ \unit{s^{-1}}$. Taking the approximation  $u_g(R)= k_{3}R$ appropriate for growing rain drops and substituting into \eqnref{eq:kernel_approx1} yields
\begin{linenomath}
\begin{subequations}
\begin{align}
    H(R,r) & =\bar{k}_r \frac{4}3\pi \rho R^3 \label{eq:kernel_approx2}   \\
    \mbox{where}\hspace{2cm}  \bar{k}_r & = 0.75 k_3/\rho \ .
    \label{eq:krbar_approx}
\end{align}
\end{subequations}
\end{linenomath}
Substituting \eqnref{eq:kernel_approx2} into \eqnref{eq:accr} shows  that the two integrals decouple and, somewhat miraculously, are both proportional to their respective masses. The end result is
\begin{linenomath}
\begin{equation}
    \dot{M}_c=-\bar{k}_r M_c M_r
    \label{eq:kr}
\end{equation}
\end{linenomath}
which is the same as \eqnref{eq:bulk_acc} except that in this estimate $\bar{k}_r$ is independent both of time and initial cloud parameters (\eqnref{eq:krbar_approx}), and evaluates to $6 \times 10^{3}\rm[cm^3/sec\cdot g]$ as utilized in \citeA{beheng_parameterization_1994}.  As described below, however, instead of using a  constant $\bar{k}_r$ we infer $\bar{k}_r$ separately for each simulation based on data from the two accretion stages, obtaining a time averaged, effective $\bar{k}_r$ (which we will show slightly varies between simulations). Note that the parameterization \eqref{eq:bulk_acc} (and variants on it) is a widely used ingredient in microphysics schemes \cite<e.g.>{beheng_parameterization_1994,ziegler_retrieval_1985,seifert_double-moment_2001}, and was validated in \citeA{zeng_two-moment_2020}.

\subsection*{Inference of the accretion parameter $\bar{k}_r$}
\label{subsec : Inferred accretion parameter}

Assuming that the cloud mass dynamics satisfies equation (\ref{eq:kr}) throughout the accretion regime, we can estimate the  cloud mass for a given $\bar{k}_r$ given $M_r(t)$ from the simulation data as 
\begin{linenomath}
\begin{equation}\label{eq:Mc_est}
    \Tilde{M}_{c}(k_{r},t)=M_{c}(t_{i})\exp[-\krbar\intop_{t_{i}}^{t}M_{r}(t')\dd t'] \ .
\end{equation} 
\end{linenomath}
An estimator for $\bar{k}_r$ can then be found by a mean least-squares fit, minimizing the difference between the measured $M_{c}$ and the expected one in \eqref{eq:Mc_est}, 
\begin{linenomath}
\begin{equation}
\partial_{k_{r}}\intop_{t_{i}}^{t_{f}}\left|\ln\Tilde{M}_{c}(\bar{k}_{r},t)-\ln M_{c}(t)\right|^{2}\dd t=0
\end{equation}
\end{linenomath}
which can be directly solved to give the formula 
\begin{linenomath}
\begin{equation}
    \bar{k}_{r}=\frac{\intop_{t_{i}}^{t_{f}}F(t)\ln \frac{M_{c}(t_i)}{M_c(t)}\dd t}{\intop_{t_i}^{t_{f}}F^{2}(t)\dd t}
\end{equation}
\end{linenomath}
where $F(t)=\intop_{t_{i}}^{t}M_{r}(t')\dd t'$. The inference is performed for each simulation separately between the initial time when $M_r/M_{0}=\Delta_{i}=0.1$ and the final time when  $M_{r}/M_{0}=\Delta_{f}=0.02$.

\begin{figure}[t!]
\noindent\includegraphics[width=1\textwidth]{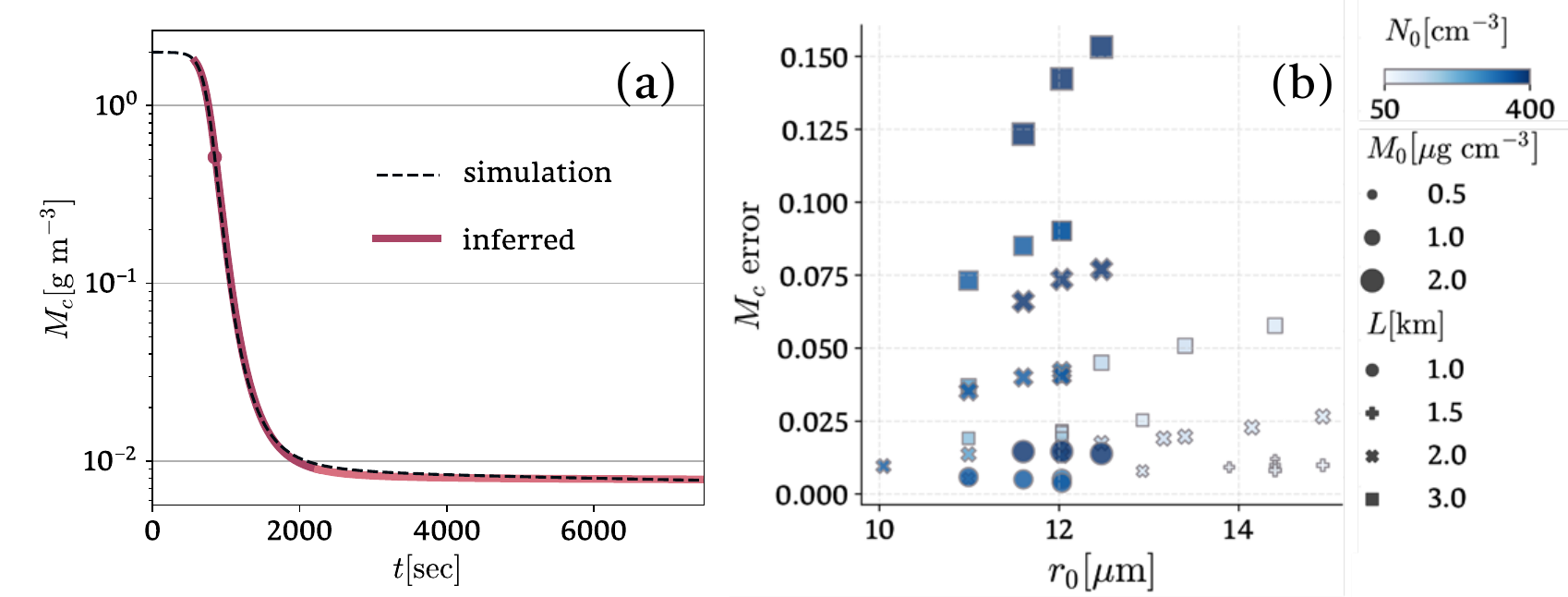}
  \caption{\emph{Evaluation of the inferred accretion rate $\bar{k}_r$ showing (a) the comparison between the measured cloud mass as a function of time (dashed line) and the expected one based on \eqref{eq:Mc_est} with the inferred $\bar{k}_r$ (red line). The duration of the fit is indicated with the dark red line while the lighter part of the curve is a continuation of the same function to a region which is not included in the fit. Simulation No. 6 in table \ref{table: simulation parameters}. (b) the relative error between the estimated cloud mass and simulated cloud mass, as defined in \eqref{eq : error of Mc}.} The inference errors are below $5\%$ for most of the simulations. The example in (a), which has a relative error $\text{of}\sim0.075$, demonstrates how larger errors are reflected in slight deviations between the measured and expected cloud mass.  
  }
  \label{fig : kr fit}
\end{figure}

Using a representative simulation, Fig. \ref{fig : kr fit}(a)  plots $M_c(t)$ as measured in the simulation together with the expected $\tilde{M}_c$ from \eqref{eq:Mc_est} with the inferred $\bar{k}_{r}$ (past the autoconversion stage). It can be seen that \eqref{eq:Mc_est} captures the dynamics well. To quantify the goodness of the fit, we also evaluate the mean squared relative error for each simulation 
\begin{linenomath}
\begin{equation}\label{eq : error of Mc}
  E =  \left(\frac{1}{t_{f}-t_{i}}\intop_{t_{i}}^{t_{f}}\left|\frac{\Tilde{M}_{c}(k_{r},t)-M_{c}(t)}{M_{c}(t)}\right|^{2}\dd t\right)^{\nicefrac{1}{2}}
\end{equation}
\end{linenomath}
and present it in \ref{fig : kr fit}(b). We see that for most simulations the error is smaller than $10\%$, and that for many of the simulations it is even lower than $3\%$.

\begin{figure}[t!]
\noindent\includegraphics[width=1\textwidth]{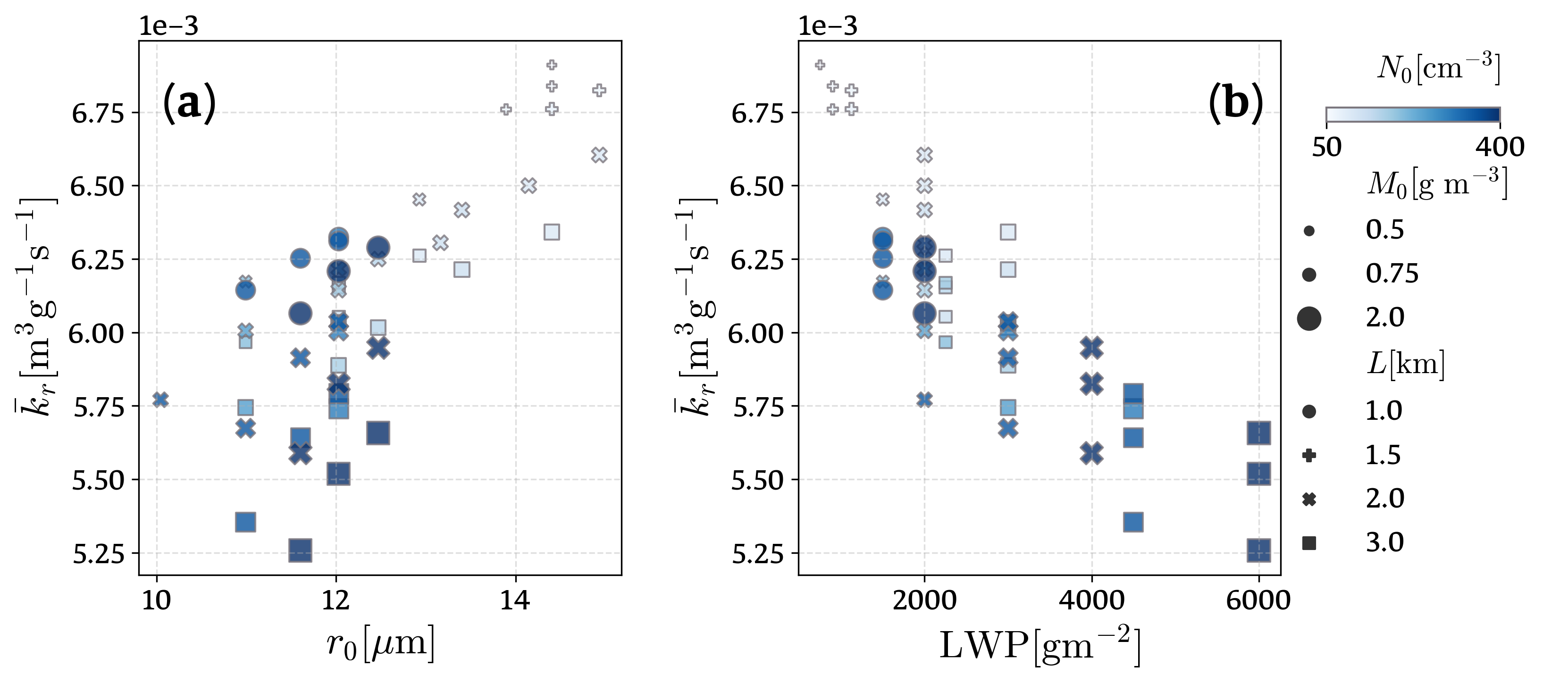}
  \caption{\textit{Inferred $\bar{k}_{r}$ values as a function of (a): $r_{0}$ and (b): LWP.} Though some trends can be observed, such as an increase with the radius and a decrease with the LWP, the data shows a spread without a clear functional relations between the parameters.}
  \label{fig : kr value}
\end{figure}

The inferred accretion parameter $\bar{k}_{r}$ for all our simulations is shown in  Fig.~\ref{fig : kr value}. While we find $\bar{k}_r\in [5.2-6.6]\times  10^{3} \ \rm [cm^3/sec\cdot g]$ which is very close to the estimate $ 0.75 k_3/\rho_{w} =6 \times 10^{3}\rm [cm^3/sec\cdot g]$, the deviations do seem to reflect physical differences between clouds. Indeed, the assumption $u_g(R)\approx k_3 R$ should not be correct throughout the evolution (cf. \eqnref{eq:ug(R)}), and the degree of error will  depend on the cloud parameters and be reflected in the effective rate $\bar{k}_r$. 
Note that instantaneously, 
\begin{linenomath}
\begin{equation}
\bar{k}_r(t) =\frac{ \intop_{40\mu m}^{\infty}  \pi R^{2}\left|u_{g}(R)\right|E(R,r_0)n(R) \dd R}{M_r}\equiv\frac{3}{4\rho}\left\langle \frac{u_g(R)}R E(R,r_0)\right\rangle_{\rm rain}
\label{eq:kr(t)}
\end{equation}
\end{linenomath}
where $\langle \cdot \rangle_{\rm rain}$ is an average with respect to the instantaneous (normalized) rain mass density function $\frac{1}{M_r(t)} 4/3 \pi \rho R^3 n(R,t)$. Thus, $\bar{k}_r(t)$ can be interpreted as an average capture rate of cloud drops by rain drops per water density $\rho$. The inferred $\bar{k}_r$ we present below captures this quantity averaged over the cloud lifetime.

Some trends are evident in Fig.  \ref{fig : kr value}. It can be seen that accretion becomes slightly more effective, i.e. $\bar{k}_{r}$ grows, with the growth of the initial drop radius $r_0$. This trend is likely inherited from the increase in the collision efficiency $E(R,r_0)$ with $r_0$, which thus increases $\bar{k}_{r}$. This dependence is evident in \eqnref{eq:kr(t)} but was neglected in our estimate \eqref{eq:krbar_approx}.
Furthermore, Fig.~\ref{fig : kr value}(b) demonstrates that $\bar{k}_{r}$ decreases with the LWP. This dependence can be explained by the fact that the rain mass for clouds with a larger LWP is able to achieve larger radii $R$, and the larger $R$ is, the smaller $u_g(R)/R$ is (see e.g. the approximate formula for the terminal velocity \eqref{eq:ug(R)}). This in turn decreases the effective $\bar{k}_r$, see \eqnref{eq:kr(t)}. Note that our inference is performed over both stages of the accretion dynamics, so that this change in $u_g(R)/R$, occurring mostly during the second stage, does influence $\bar{k}_{r}$ (In particular, we get different results for $\bar{k}_r$ if we infer it separately before and after $M_r$ peaks)  .

\subsection*{The effect of the inferred $\bar{k}_{r}$ on the collapse}
\label{subsec : static mu collapse}

\begin{figure}[t!]
     \includegraphics[width=1\textwidth]{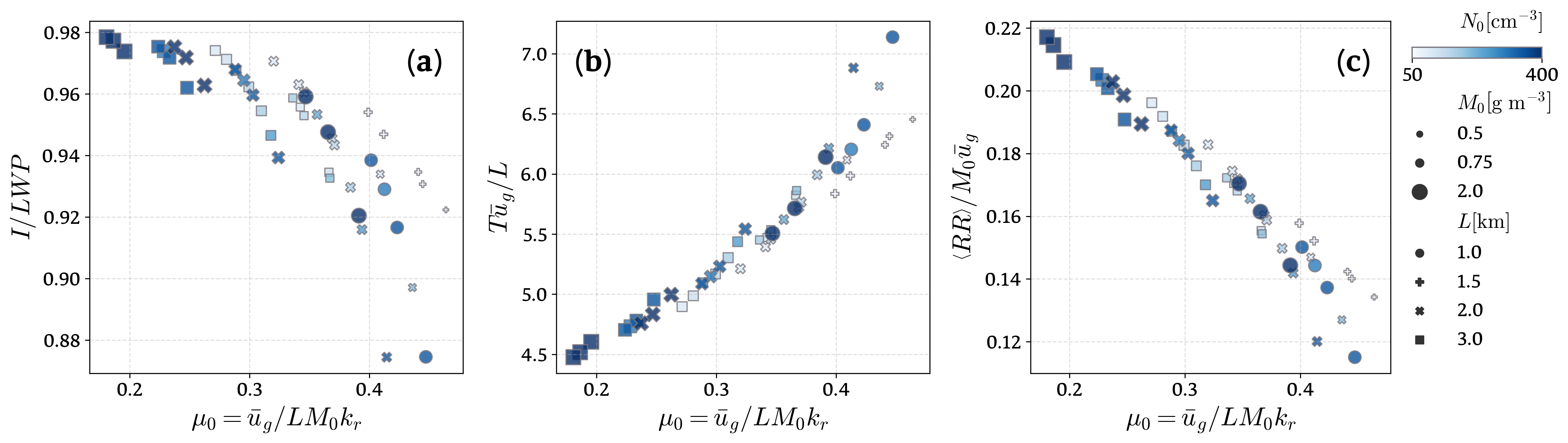}
     \caption{\emph{Deterioration in the collapse of cloud observables when plotted as a function of $\mu_0$, where the inferred $\bar{k}_r$ is replaced by a fixed value $k_r$ as seen for the (a) normalized accumulated rain (b) the lifetime in fall-out units, $T\bar{u}_{g}/L$ and (c) the normalized mean rain rate.} }
     \label{fig: I T and RR vs mu0}
 \end{figure}

The variations we find in the inferred $\bar{k}_r$ between different simulations may appear insignificant, and one might expect that using a constant $k_r$ for all the clouds should work equally well. Therefore, here we demonstrate that when using $\bar{u}_g$ for the time normalization and definition of $\mu$, but replacing $\bar{k}_r$ by $k_r=6 \unit{m^3\ kg^{-1} s^{-1}}$, as discussed in section \ref{ssec: assum and form} and equation \ref{eq:krbar_approx},
the collapse of the various cloud observables deteriorates.  
Figure \ref{fig: I T and RR vs mu0} shows the three cloud observables as a function of $\mu_{0}$ which is defined like $\mu$ but with a constant value of $\bar{k}_r$.

\section{Inference of the effective sedimentation speed} \label{appendix_sed}

For the inference of the effective sedimentation speed $\bar{u}_g$ we use a similar approach to that used for $\bar{k}_r$. Indeed, during the rainfall stage, the expected accumulated rain mass that fell out, $W(t)$, given the in-cloud rain mass $M_r(t)$ is given in our bulk dynamic formulation \eqref{eq:rainfall} by  
\begin{linenomath}
\begin{equation}
   \Tilde{W}(\ugbar,t)=\frac{\ugbar}{L}\intop_{t^*}^{t}M_{r}(t')\dd  t'+W(t^*)
   \label{eq: Z_est}
\end{equation}
\end{linenomath}
where $t^*$ is the dimensionful time corresponding to \taustar\ defined in \eqnref{eq:taustar_def}, well within the rainfall regime.

By using a mean least-squares fit to minimize the difference between \eqref{eq: Z_est} and the measured $W(t)=M_{0}-M_{c}-M_{r}$ we find an estimator for $\bar{u}_g$: 
\begin{linenomath}
\begin{equation}
\partial_{\ugbar}\intop_{t^*}^{t_{f}}\left|\Tilde{W}(t)-W(t))\right|^{2}\dd t=0
\end{equation}
\end{linenomath}
which gives
\begin{linenomath}
\begin{equation}
    \frac{\bar{u}_{g}}{L}=\frac{\intop_{t_{f.}}^{t_{f}}(W(t)-W(t^*))F(t)\dd t}{\intop_{t^*}^{t_{f}}F^{2}(t)\dd t}
\end{equation}
\end{linenomath}
where $F(t)=\intop_{t^*}^{t}M_{r}(t')\dd t'$. The final time for the fit, $t_f$, is the same as used throughout the text, defined via $M_{r}(t_f)=\Delta_{f}M_{0}$. 

\begin{figure}[t!]
  \noindent\includegraphics[width=1\textwidth]{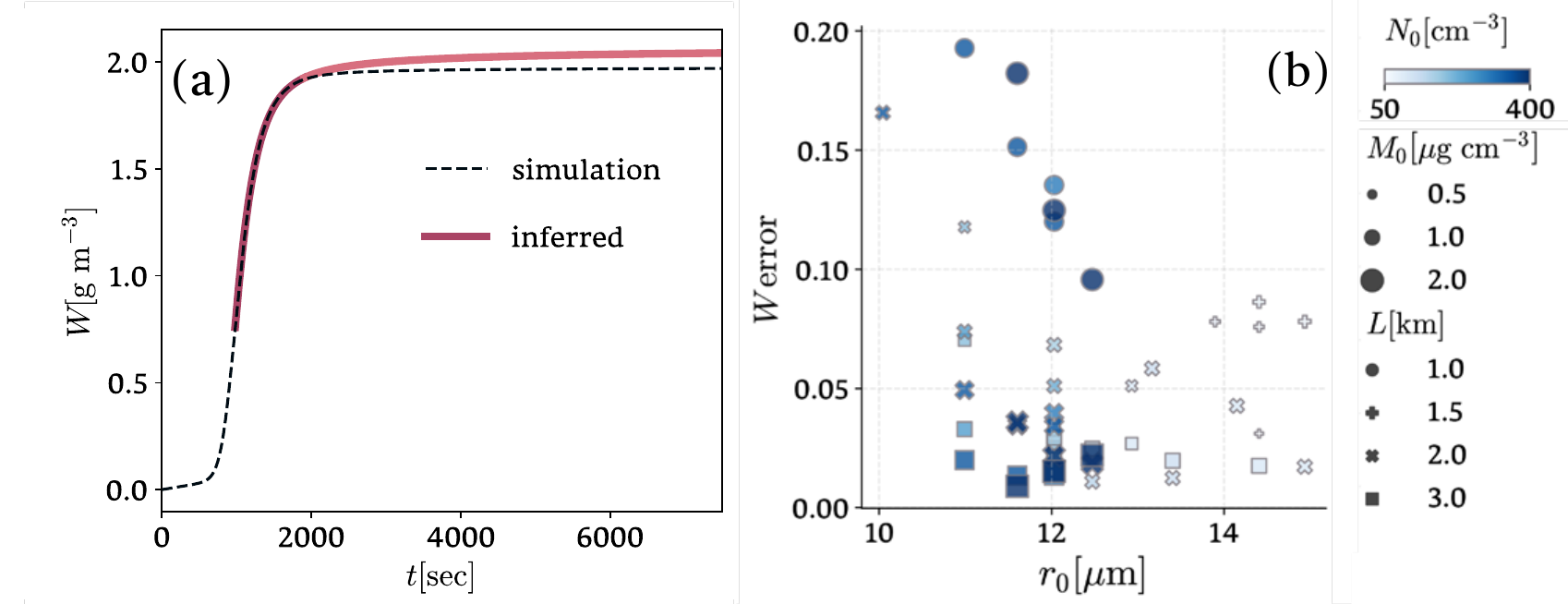}
  \caption{\emph{Evaluation of the inferred sedimentation speed $\bar{u}_g$ showing (a) a comparison between the measured accumulated rain mass as a function of time (dashed line) and the estimated one based on Eq. \eqref{eq: Z_est}, with the inferred $\bar{u}_g$ (red line). As in \ref{fig : kr fit} the duration of the fit is indicated with the darker red line and the lighter one is a continuation of the same function. Simulation No. 31 in table \ref{table: simulation parameters}. (b) the relative error between the measured accumulated rain mass and the estimated one, as defined in \eqref{eq : error of Mr}.} The inference errors are below $ 7\%$ for most of the simulations. The example shown in (a) has a larger relative error than the typical ones: $\text{equal to}\sim0.125$. For such simulations the dynamics are still well represented during the inference period, but an overestimation of the final accumulated rain mass can be seen.}
  \label{fig : ug fit}
\end{figure}

\begin{figure}[t]
\noindent\includegraphics[width=1\textwidth]{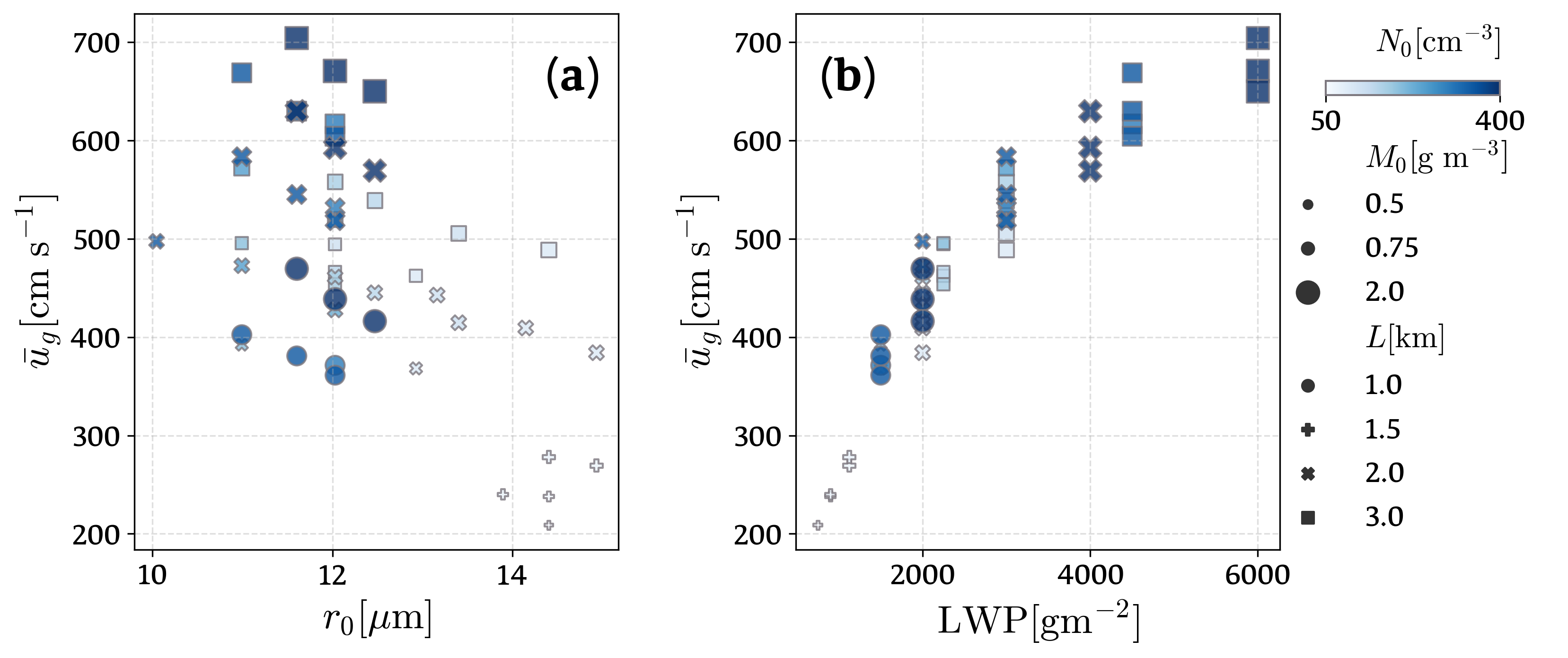}
  \caption{\textit{Inferred values of the sedimentation speed, $\bar{u}_{g}$,  as a function of (a): $r_{0}$ and (b): LWP.} While there is a clear correlation between the effective sedimentation speed and the LWP, for a fixed LWP, a spread of the order of $\approx 100 \unit{cm /s}$ in $\bar{u}_{g}$ still remains.}
  \label{fig : ug value}
\end{figure}

In Fig. \ref{fig : ug fit}(a), we demonstrate how well the expected accumulated rain mass \eqref{eq: Z_est}, with the inferred $\bar{u}_{g}$, captures the rainfall dynamics in the interval over which the fit is performed. We quantify the goodness of the fit of the effective dynamics using the relative error estimate defined by
\begin{linenomath}
\begin{equation}\label{eq : error of Mr}
  E =  \left(\frac{1}{t_{f}-t^*}\intop_{t^*}^{t_{f}}\left|\frac{\Tilde{W}(\bar{u}_{g},t)-W_{r}(t)}{W_{r}(t)}\right|^{2}\dd t\right)^{\nicefrac{1}{2}}.
\end{equation}
\end{linenomath}
The results are presented in Fig. \ref{fig : ug fit}(b), showing that for most of our simulations the error is less than $10\%$.   
The finite gap between the inferred $W$ and the measured one at the long time limit is typical in our results and has to do with the temporal development of the bulk fall speed, section \ref{sec_sed}. 
We present the inferred terminal speed $\bar{u}_g$ for the clouds in our simulations in Fig.~\ref{fig : ug value}. The inferred terminal speeds we find are all in the range $[200-400]\unit{cm/s}$, roughly corresponding to raindrops of radii $[0.2-1]\unit{mm}$. Also note that clouds which have the smallest terminal speeds $\sim 200 \unit{cm/s}$ do not have an extreme value of $\mu$, as the small terminal speed is compensated by (and originates from) a low LWP (Fig.~\ref{fig : ug value}(b)).

We now comment on the dependence of $\bar{u}_g$ on the initial bulk parameters. As seen in Fig.~\ref{fig : ug value}(a), $\bar{u}_g$ does not have any visible trend with $r_0$, but strongly depends on the LWP, as expected. Indeed, the larger the LWP is, the larger the radii that the rain drop population can attain, and so also the terminal speed. Thus, also the effective terminal speed $\bar{u}_g$ for such a cloud increases, see e.g. Eq.~\eqref{def:avg ug}.  

\section{Detailed analytical formulae for cloud observables}
\label{app: analytical approx}
\subsection{Accumulated rain}
\label{app:accumulated_rain}
Here we expand upon the derivation of the analytical formulae shown in the main text, starting with the accumulated rain.
Equation \eqref{eq:mr_w_star} is an exact formula for the final accumulated rain mass within the scope of our effective model. Rewriting this equation with $m_{r}(\tau_{f})=\Delta_{f}$ and choosing a time $\tau_r$ such that  $w(\tau_r)\approx 0$ we obtain 
\begin{linenomath}
\begin{equation}
w(\tau_{f})=1-\Delta_{f}-m_{c}(\tau_r)e^{-w(\tau_{f})/\mu}
\label{eq: final w formula}
    \end{equation}
\end{linenomath}
 The choice of the time $\tau_r$ will be discussed in more detail below, here we just comment that it should be chosen such that the rainfall regime already applies but that the accumulated rain mass is still small. Also recall that our model has two regimes, and in the mass conserving regime certainly $w(\tau)\approx 0$ so a reasonable choice for $\tau_r$ is the end of the mass conserving regime. 

Equation \ref{eq: final w formula} is a closed equation for the final accumulated mass $w(\tau_f)$ which can be expressed in terms of the Lambert $W$ function, defined as the solution $y=W_0(x)$ to the equation $y=xe^{-y}$, where $x>-e^{-1}$. Explicitly
\begin{linenomath}
\begin{equation}
    \frac{\text{I}}{\text{LWP}}=w(\tau_f)=1-\Delta_f +\mu W_0\left(- \frac{m_{c}(\tau_r)}{\mu} e^{\frac{-1+\Delta_f}\mu}\right)
    \label{eq:I_formula_exact}
\end{equation}
\end{linenomath}
giving an exact solution of equations \ref{eq:mass_conserving_re} with the initial condition taken at time $\tau_r$ assuming $w(\tau_r)=0$. 
To obtain a more explicit formula we take the leading order in the expansion of the Lambert $W$ function, $W_{0}(x)\approx x$, and use that  $\Delta_{f}\ll 1$, which gives
\begin{linenomath}
\begin{equation}
   w(\tau_{f})\approx 1-\Delta_{f}-m_{c}(\tau_r) e^{\frac{-1+\Delta_{f}}{\mu}}\approx 1-\Delta_{f}-m_{c}(\tau_r) e^{-1/\mu}
    \label{eq: w_f approx}
\end{equation}
\end{linenomath}
The final result presented in the main text is then achieved by choosing the transition time $\tau_r$ at the peak of $m_{r}$ so that $m_{c}(\tau_r)=\mu$.

\subsection{Cloud lifetime}
\label{app:cloud_lifetime}
The cloud lifetime in the model is determined by adding the duration  of the in-cloud mass conserving stage and that of the fall-out stage. The duration of the fall-out stage is given by
\begin{linenomath}
\begin{equation}
  \tau_f-\tau_r=\int_{0}^{w(\tau_f)}\frac{\dd w}{\mu\left(1-w-m_{c}(\tau_r) e^{-w/\mu}\right)} 
  \label{eq:time}
\end{equation}
\end{linenomath}
while the duration of the in-cloud mass conserving regime is
\begin{linenomath}
\begin{equation}
\tau_r-\tau_{0}=\ln\frac{(1-\Delta_{i})(1-m_{c}(\tau_r))}{m_{c}(\tau_r) \Delta_{i}}
\label{eq:mass_cons_time}
\end{equation}
\end{linenomath}
which is obtained by setting $m_r(\tau_0)=\Delta_i$ and $m_c(\tau_0)=1-\Delta_i$ in \eqref{subeq : mass conserving solutions Mc}. Here we are assuming that the two regimes of the model can be directly connected. The transition time is denoted by $\tau_r$, and we assume that $w(\tau_r)=0$ as it is preceded by the mass conserving regime. In total we have the prediction
\begin{linenomath}
\begin{equation}
T\frac{\bar{u}_g}{L}=\mu (\tau_f -\tau_{0})=\mu \ln\frac{(1-\Delta_{i})(1-m_{c}(\tau_r))}{m_{c}(\tau_r) \Delta_{i}}+\int_{0}^{w(\tau_f)}\frac{\dd w}{\left(1-w-m_{c}(\tau_r) e^{-w/\mu}\right)} 
\label{eq: normalized lifetime_exact}
\end{equation}
\end{linenomath}

We now wish to obtain a closed formula for the duration of the rainfall stage, $\tau_f-\tau_r$, Eq. \eqref{eq:time}.  
We present a more detailed derivation as compared to the main text, emphasizing why the approximation is expected to capture the main contribution when $\Delta_f$ is small. 
Indeed, to approximate the integral \eqref{eq:time}, we notice that in the long time limit all the rain mass falls out, and  since
\begin{linenomath}
\begin{align}\label{eq: linear approx of f}
    \left(1-w-m_c(\tau_r)e^{-w/\mu}\right)= m_r(\tau) 
\end{align}
\end{linenomath}
the integral \eqref{eq:time} diverges logarithmically. From \eqref{eq:mr_w_star}, this happens at the value $w_{\infty}$ which satisfies 
\begin{linenomath}
    \begin{equation}
        0=1-w_{\infty}-m_c(\tau_r) e^{-w_{\infty}/\mu}
    \end{equation}
\end{linenomath}
Since $\Delta_f\ll1$, we expect $w(\tau_f)\sim w_{\infty}$ and for the lifetime to be dominated by the region close to $w(\tau_f)$. Thus, we shall approximate the denominator around $w_{\infty}$:
\begin{linenomath}
\begin{align}
    \mu\left(1-w-m_c(\tau_r)e^{-w/\mu}\right)\approx - (\mu-m_c(\tau_r)e^{-w_{\infty}/\mu})(w-w_{\infty})
\end{align}
\end{linenomath}
which gives the relation
\begin{linenomath}
\begin{align}
    \Delta_f\approx - \left(1-\frac{m_c(\tau_r)}{\mu}e^{-w_{\infty}/\mu}\right)(w(\tau_f)-w_{\infty})=w_\infty (w_\infty-w(\tau_f))
\end{align}
\end{linenomath}
The fall-out time now reads \eqref{eq:time}
\begin{linenomath}
    \begin{align}
        \tau_f-\tau_r\approx -\intop_{0}^{w(\tau_f)}\frac{\dd w}{((\mu-m_c(\tau_r)e^{-w_{\infty}/\mu})(w-w_{\infty}))}=\frac{1}{(\mu-m_c(\tau_r)e^{-w_{\infty}/\mu})}\ln\frac{w_{\infty}}{w_{\infty}-w(\tau_f)}
        \label{eq:w_dif}
    \end{align}
\end{linenomath}
Now, since we have already made a few rather crude approximations, we will only take the leading order approximations for $w_\infty\approx 1$. Indeed, the corrections to these results are smaller than the errors we've made in our approximations so far.   
Using \eqref{eq:w_dif} we find
\begin{linenomath}
\begin{equation}
  \tau_f-\tau_r\approx  \frac{1}{(\mu-m_c(\tau_r)e^{-\frac{1}{\mu}})}\ln\left[\frac{1}{\Delta_{f}}\right] =\frac{1}{\mu }\frac{\ln\left[\frac{1}{\Delta_{f}}\right]}{(1-\frac{m_{c}(\tau_r)}\mu e^{-\frac{1}{\mu}})} 
  \label{eq: approx lifetime}
\end{equation}
\end{linenomath}
We have left $m_{c}(\tau_r)$ as a free parameter here, and we will discuss its choice in the next section.

There are several things to note regarding formula \eqref{eq: approx lifetime}. First, in terms of the dependence on $\mu$, we can read the relevant fall-out time scale from the right hand side of \eqref{eq: approx lifetime}: instead of the rate $\mu$ one has $ \mu(1-(m_c(\tau_r)/\mu)e^{-\frac{1}{\mu}}))$, meaning that there is an effective fall-out rate which is slightly smaller than the original $\nicefrac{\bar{u}_{g}}{L}$. 
This smaller effective rate can be attributed to the fact the while the rain drop mass reduces due to sedimentation, it also gains mass due to the accretion of small cloud drops. This results in a smaller effective fall-out rate. In particular, we can interpret this effective rate to be $\mu -m_c^{\infty}$, meaning that $m_c$ is kept fixed to its final value $m_c^{\infty}\approx m_c(\tau_r) e^{-\frac{1}{\mu}}$, causing $m_r$ to decay with this effective rate in \eqref{eq:system}. This is equivalent to an assumption that the rain mass continues to evolve, while the cloud mass saturates to its final value. Thus, since the larger $\mu$ is, the larger the cloud mass that remains, this gives rise to a slower decay of the rain mass, as we indeed see in Fig.~\ref{fig: accumulated rain mass normalized}. Also note that it is only the rain mass which has its decay rate altered, and the increase in the accumulated rain mass still happens with the sedimentation rate $\mu$. Thus, the appearance of the effective rate in the lifetime reflects that it is related to the amount of the in-cloud rain mass. 

Finally, note the form of the dependence of the lifetime on the final rain mass $\Delta_f$, which is of the form $-\ln\Delta_f$ due to the logarithmic divergence of the time to reach zero rain mass, as we mentioned above. 

For completeness, we write the resulting formula for the lifetime with the transition time between regimes chosen at $m_c=m_c(\tau_r)$
\begin{linenomath}
\begin{equation}
T\frac{\bar{u}_g}{L}=\mu (\tau_f -\tau_{0})=\mu \ln\frac{(1-\Delta_{i})(1-m_{c}(\tau_r))}{m_{c}(\tau_r) \Delta_{i}}+\frac{1}{(\mu-m_{c}(\tau_r)e^{-\frac{1}{\mu}})}\ln\left[\frac{1}{\Delta_{f}}\right]
\label{eq:T_tot_1}
\end{equation}
\end{linenomath}
which gives \eqref{eq: normalized lifetime} in the main text when one uses $m_c(\tau_r)=\mu$.
\subsection{The transition time}
\label{app: transition time}
As mentioned above, our formulae for the observables contain an undetermined parameter, $m_c(\tau_r)$, where $\tau_r$ is the time of transition between the mass conserving and rainfall regimes in our modeling, assumed to be sharp in order to connect the two regimes directly. In the main text we have further assumed that $\tau_r=\taupeak$ which gives $m_c(\tau_r)=\mu$. 
 Section \ref{sec_sed} emphasizes the time-dependent nature of the fall-out velocity, suggesting that the transition between the temporal regimes should be represented as continuous. Using an abrupt transition instead is expected to be valid when the timescale of the change in fallout speed is significantly shorter than the lifetime, or at least shorter than the time during which most of the changes occur.
 Although Figure \ref{fig : ug dynamics} doesn't seem consistent with the former assumption, the latter is consistent with the observed dynamics, as discussed in the main text. Thus, the fundamental physical properties of the dynamics seem to be captured by our simplified model.

\begin{figure}[t]
\noindent\includegraphics[width=1\textwidth]{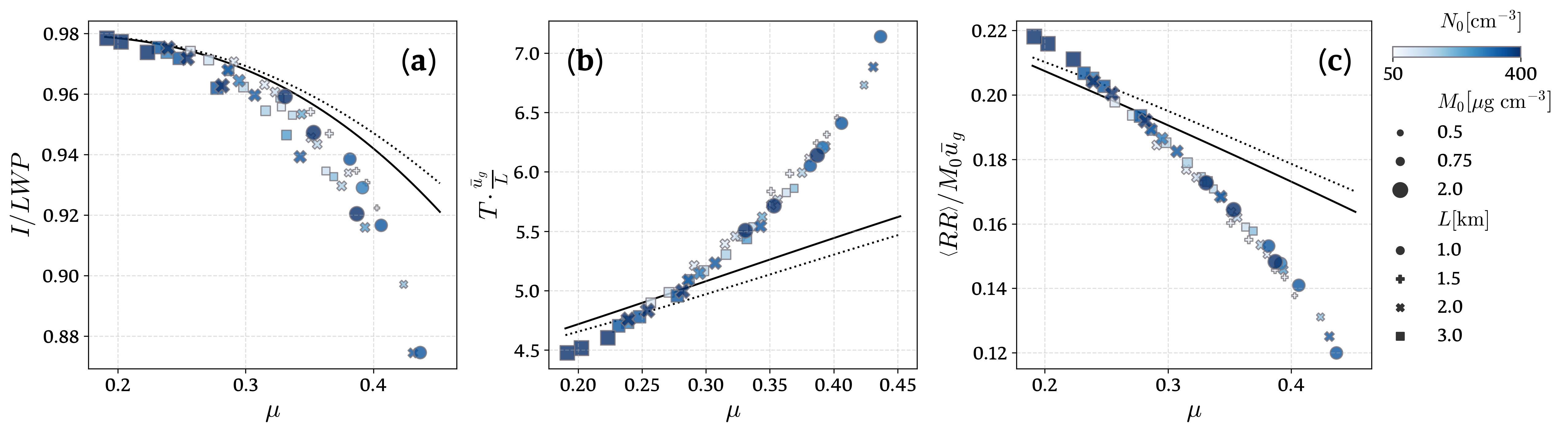}
  \caption{\emph{Normalized cloud observables compared to results from the model, the transition time $\tau_r$ between the mass conserving and rainfall regime is taken to be such that $ m_c(\tau_r)=\mu$. Exact numerical solutions of the model in black solid lines: using Equation \eqref{eq:I_formula_exact} for the normalized accumulated rain and Equation \eqref{eq: normalized lifetime_exact} for the lifetime. The approximate formulae are shown in dotted grey lines: Equation \eqref{eq:wf} is used for the normalized accumulated rain and \eqref{eq: normalized lifetime} for the lifetime.} While the approximate formulae are close to the exact results for the model, there is a visible discrepancy between the degree of variation with $\mu$ and functional dependence between model and simulation results.}
  \label{fig : results s=1}
\end{figure}

Let us now discuss the best choice for the transition time $\tau_r$. In 
Fig. \ref{fig : results s=1} we present the cloud observables as a function of $\mu$ alongside the predictions from the model, choosing $m_c(\tau_r)=\mu$ as in the main text. The black solid line shows the exact solution of \ref{eq:w_integral_star} and \ref{eq:I_formula_exact}, and the grey dotted line shows the analytical formulae presented above, the two being very similar. 
The model results with this choice of $\tau_r$  are able to capture the characteristic values of the observables, as well as their trends, but do not seem to capture the observed functional dependence. In particular, the subtle increase of the lifetime with $\mu$ in the model is linear and is very slight, due to the contribution from the mass conserving portion of the dynamics. In particular,  the correction to the duration of the rainfall regime as compared to $1/\mu$ discussed in the previous section is in fact negligible.  

\begin{figure}[t]
\noindent\includegraphics[width=1\textwidth]{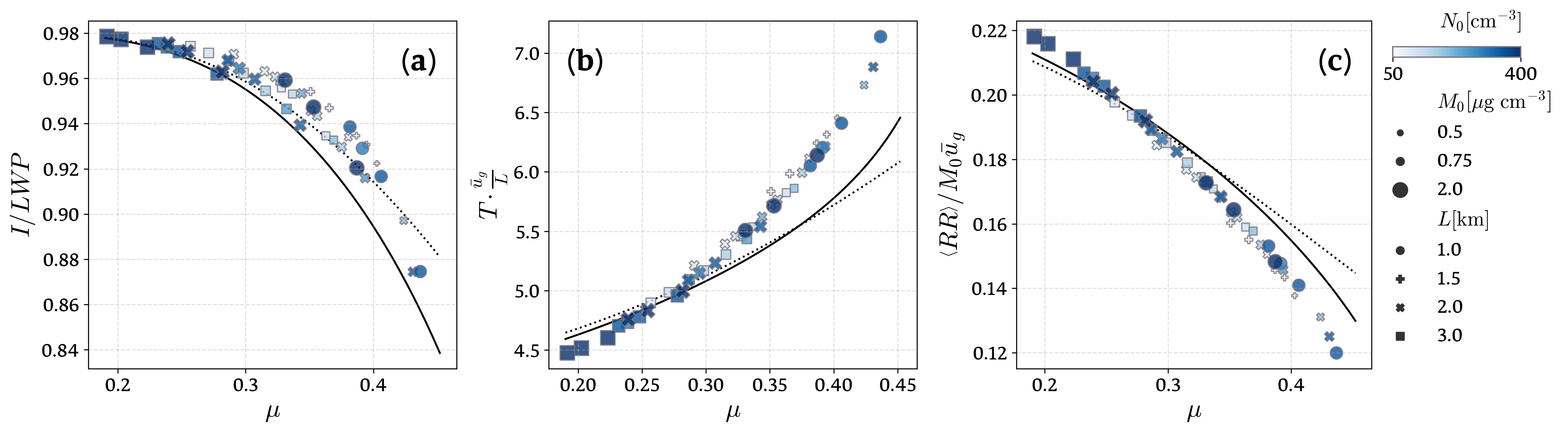}
  \caption{emph{Normalized cloud observables compared to results from the model, the transition time $\tau_r$ between the mass conserving and rainfall regime is taken to be such that $ m_c(\tau_r)=2\mu$. Exact numerical solutions of the model in black solid lines: using Equation \eqref{eq:I_formula_exact} for the normalized accumulated rain and Equation \eqref{eq: normalized lifetime_exact} for the lifetime. The approximate formulae are shown in dotted grey lines: Equation \eqref{eq: w_f approx} is used for the normalized accumulated rain and \eqref{eq:lifetime_2} for the lifetime.} The approximate formulae are close to the exact results for the model, and are also able to capture the main variation with $\mu$, though not the exact functional dependence seen in simulations. }
  \label{fig : results s=2}
\end{figure}

One possible source for the disparity is the choice of the transition time to be when $m_r$ is maximized. Instead, we might expect that the transition time should be taken before the peak in $m_r$ is reached. Indeed, figure \ref{fig: dynamics} shows that larger than $10\%$ deviations from the predictions of the mass conserving model develop before the peak of $m_{r}$ is reached, for all the simulations shown. Also notice that the value of $m_{r}$ never reaches $1-\mu$, the maximal value of $m_{r}$ assuming the mass conserving evolution up to the peak, indicating the loss of in-cloud mass before that. This motivates choosing an earlier time for the transition, and we choose $m_c(\tau_r)=2\mu$, keeping the linear dependence on $\mu$ so that the transition will happen at a higher value of $m_c$ (and thus earlier) for higher $\mu$. The coefficient $2$ is chosen empirically to produce a good match between the prediction for $I/LWP$ as a function of $\mu$ and that obtained in the simulations, as shown in figure \ref{fig : results s=2}$(a)$ in the solid line. Note that for $\mu=0.5$ this choice gives $\tau_r=0$ implying that the cloud is in the rainfall regime already when $m_r=\Delta_i=0.1$. Figures \ref{fig : results s=2}$(b),(c)$ further show the comparison between the predictions from the model and the other two observables. The results from the model, plotted with a solid black line, indeed do a better job at capturing the more subtle trends with $\mu$ as compared to our previous choice for $\tau_r$, Fig. \ref{fig : results s=1}. Still, the functional form of, e.g. the lifetime, is not fully captured by this choice.

Choosing an earlier time for the transition also implies slightly different approximate formulae for the observables, shown as dashed lines in figure \ref{fig : results s=2}. However, given that our approximations for the lifetime leading to Eq. \ref{eq: approx lifetime} were rather crude, this formula cannot be directly used. Indeed, in deriving this formula we have assumed that the dynamics are dominated by late times, when the cloud mass practically reaches its final value, which cannot be a good representation of the dynamics before the peak in $m_r$, when the dynamics of the rain mass are domintaed by accretion. Instead, we will approximate the duration of the time-lag $\tau_r-\taupeak$ by what would be expected from the mass conserving dynamics, using equation \eqref{eq:mass_cons_time} up to $\taupeak$,
\begin{linenomath}
\begin{equation}
 \tau_f -\taupeak= \ln\frac{(1-\Delta_{i})(1-\mu)}{\mu \Delta_{i}} 
\end{equation}
\end{linenomath}
 The lifetime then reads 
\begin{linenomath}
\begin{equation}
\tau_f -\tau_{0}=\ln\frac{(1-\Delta_{i})(1-\mu)}{\mu \Delta_{i}} +\int_{w(\taupeak)}^{w(\tau_f)}\frac{\dd w}{\left(1-w-m_{c}(\tau_r) e^{-w/\mu}\right)} 
\end{equation}
\end{linenomath}

Next, using equation \eqref{eq:mc_formula} we get that the accumulated rain up to the peak is given by
\begin{linenomath}
\begin{equation}
w(\taupeak)=\mu \ln\left[\frac{m_c(\tau_r)}{m_c(\taupeak)}\right]=\mu \ln2 \ll1
\end{equation}
\end{linenomath}
Assuming that the integral on the right hand side is dominated by the long time region as we have above, we get 
\begin{linenomath}
    \begin{align}
        \tau_f-\taupeak \approx \frac{1}{(\mu-m_c(\tau_r)e^{-w_{\infty}/\mu})}\ln\frac{w_{\infty}-w(\taupeak)}{w_{\infty}-w(\tau_f)}
    \end{align}
\end{linenomath}
Finally, since $w(\taupeak)\ll w_{\infty}$ we can neglect it and arrive at the same final result we had above, 
\begin{linenomath}
\begin{equation}
  \tau_f-\taupeak\approx  \frac{1}{(\mu-m_{c}(\tau_r)e^{-\frac{1}{\mu}})}\ln\left[\frac{1}{\Delta_{f}}\right] 
\end{equation}
\end{linenomath}
where we should now plug in $m_c(\tau_r)=2\mu$ to get
\begin{linenomath}
\begin{equation}
  T\frac{\bar{u}_g}L=\mu(\tau_f-\tau_0)=  \mu\ln\frac{(1-\Delta_{i})(1-\mu)}{\mu \Delta_{i}}+ \frac{1}{(1-2e^{-\frac{1}{\mu}})}\ln\left[\frac{1}{\Delta_{f}}\right] 
  \label{eq:lifetime_2}
\end{equation}
\end{linenomath}

We mention that the integral in Eq. \eqref{eq:w_integral_star} can be approximated analytically in a more detailed way by segmenting the fall-out regime into e.g. three sub-regimes, expanding around the corresponding value of $w$ in each regime, but the resulting analytical formulae are involved and do not provide additional insights.

\section{Comparison of the dynamics}
\label{app: comparison of the dynamics}
For completeness here we compare the bulk dynamics in the model to those in the simulations, going beyond the averaged quantities discussed above. The form of the bulk equations in the two separate temporal regimes was discussed both in Sec. \ref{sec_bulk_solution} of the main text and in \ref{appendix_acc}-\ref{appendix_sed}. Separately, the equations seem to capture the dynamics well, though the time over which each regime is valid varies between the simulations, mainly dictated by the value of $\mu$. 
As for the observables discussed in app.\ref{app: transition time}, in our bulk modelling we use the assumption of a sharp transition between the two regimes, taken to be at the time when $m_c(\tau_r)=2\mu$.
The solutions for the cloud and rain mass as a function of time are then given by  Eq. \eqref{eq : mass conserving solutions} in the mass conserving regime, and can be read from the exact relations Eqs.\eqref{eq:time}, \eqref{eq:mc_formula}, \eqref{eq:mr_w_star}, which can be computed numerically, in the rainfall regime.

\begin{figure}[t]
\noindent\includegraphics[width=0.8\textwidth]{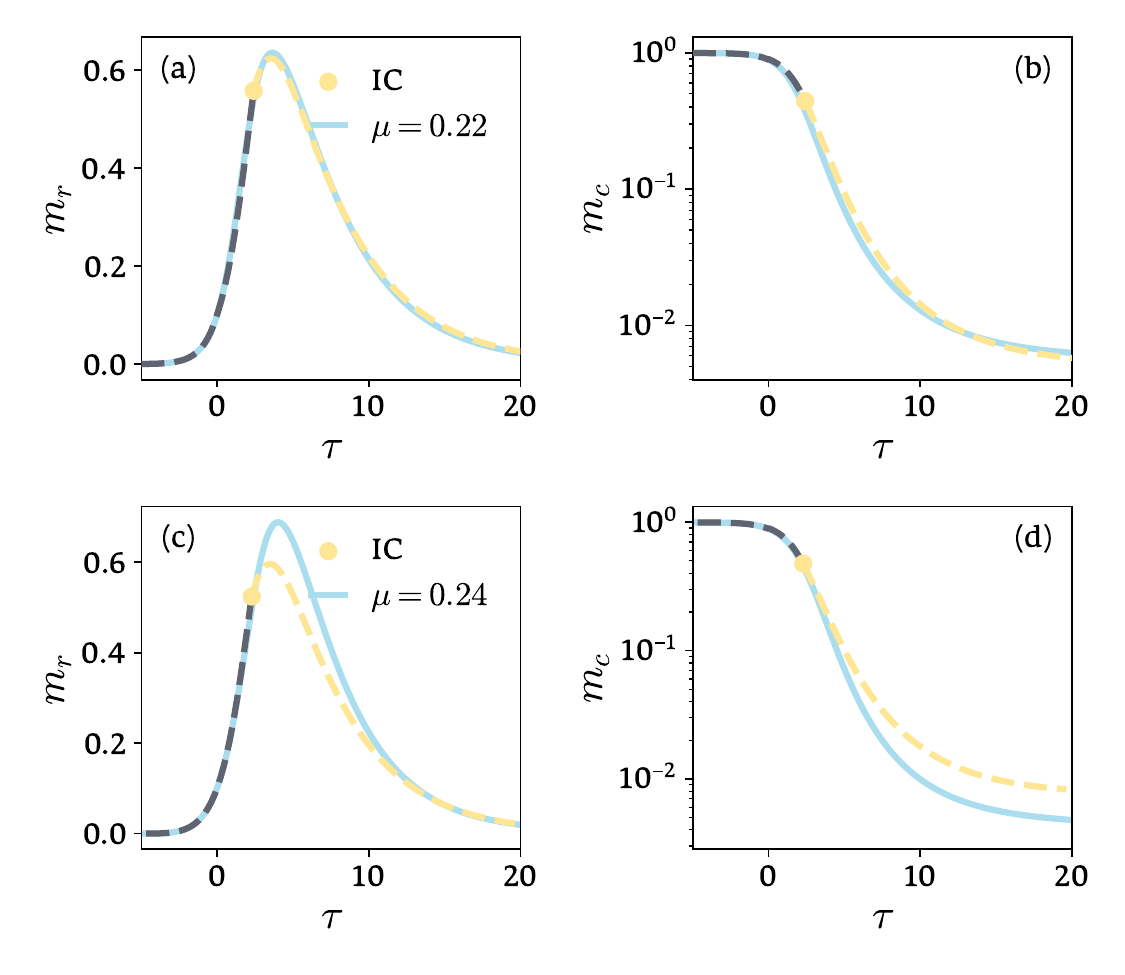}
  \caption{\emph{Comparison of the bulk dynamics between the model (dashed lines) and two simulations, simulations no. 3 and 6 in table \ref{table: simulation parameters} (solid light blue line) with a similar value of $\mu$. The normalized rain mass dynamics is shown in the left column and the cloud mass dynamics in the right column. The upper row corresponds to $\mu=0.22$, the lower row to  $\mu = 0.24$. The grey line shows the mass conserving solutions, Eq. \eqref{eq : mass conserving solutions}, and the orange line the fall-out solution, Eqs.\eqref{eq:time}, \eqref{eq:mc_formula}, \eqref{eq:mr_w_star}. The transition time between regimes $\tau_r$ is chosen so that $m_{c}(\tau_r)=2\mu$ and is marked with a circle.}  }
  \label{fig: dynamics comp small}
\end{figure}

\begin{figure}[t]
\noindent\includegraphics[width=0.8\textwidth]{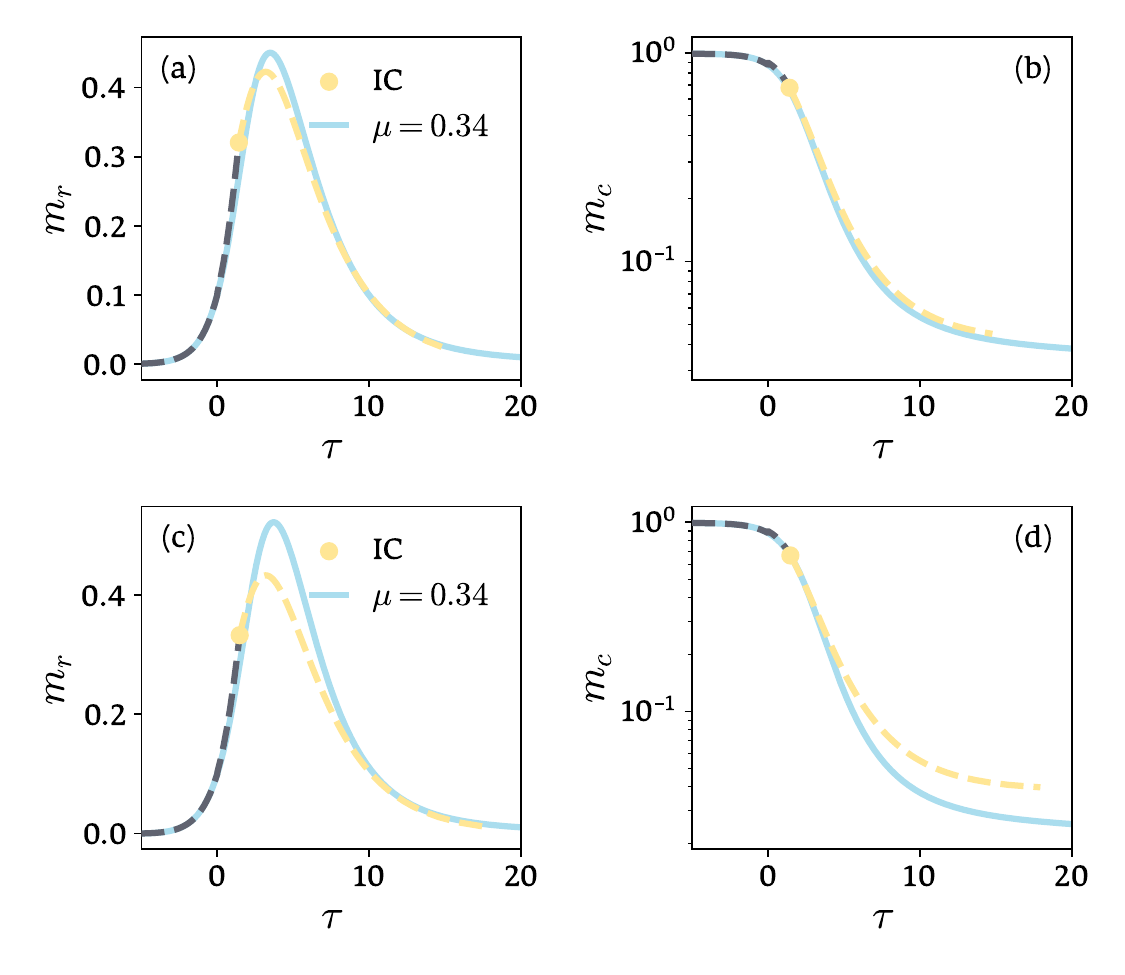}
  \caption{\emph{\emph{Comparison of the bulk dynamics between the model (dashed lines) and two simulations (solid light blue line) with a similar value of $\mu$. As in figure \ref{fig: dynamics comp small} but with $\mu=0.34$. } Simulations numbers 26 and 27 in table \ref{table: simulation parameters}}}
  \label{fig: dynamics comp mid}
\end{figure}

\begin{figure}[t]
\noindent\includegraphics[width=0.8\textwidth]{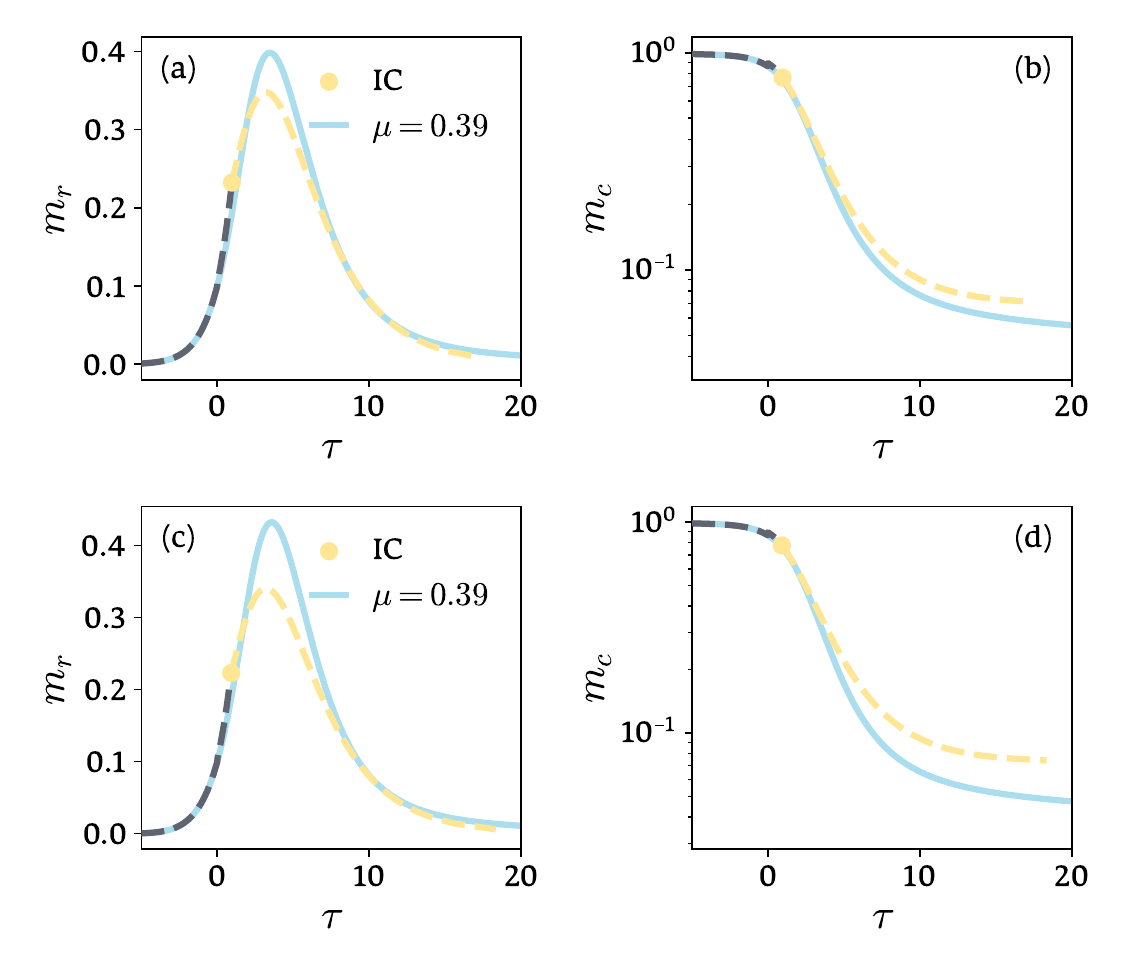}
  \caption{\emph{Comparison of the bulk dynamics between the model (dashed lines) and two simulations (solid light blue line) with a similar value of $\mu$. As in figure \ref{fig: dynamics comp small} but with $\mu=0.39$. Simulations numbers 40 and 41 in table \ref{table: simulation parameters}} }
  \label{fig: dynamics comp large}
\end{figure}

In figures \ref{fig: dynamics comp small}-\ref{fig: dynamics comp large} we compare the rain and cloud mass dynamics to solutions of the model with the corresponding $\mu$ for six different clouds.
We have chosen the examples such that they also demonstrate the degree of variance in the dynamics between simulations with similar inferred values of $\mu$. Clouds with the same $\mu$ values should in principle follow the exact same dynamics, and so compare to the model in the same way. For the most part this is indeed what we observe, yet there are outliers throughout the range of values of $\mu$ as demonstrated in \ref{fig: dynamics comp small}-\ref{fig: dynamics comp large}.
In particular, we have chosen the examples in figures \ref{fig: dynamics comp small}-\ref{fig: dynamics comp large} such that for each range of $\mu$ values there is one example representative of cases with the best match to the model dynamics, while the other is representative of the worst one.  


Overall, these six examples demonstrate that our modeling works better for smaller values of $\mu$. 
Recall that small $\mu$ implies a larger separation between accretion and sedimentation timescales. Thus, we expect a relatively shorter time when the two rates are of the same order.
Thus, for smaller values of $\mu$ the time over which the momentary sedimentation speed $\bar{u}_g(\tau)$ attains the value $\bar{u}_g$ (reflected in $\mu$) is shorter (Fig.\ref{fig : ug dynamics}), so that we expect that neglecting this phase of the dynamics incurs smaller errors.  Indeed, the main discrepancy between the model and simulation bulk dynamics arises from the temporal region between $\tau_r$ and when the rain mass is maximized, during which the effective sedimentation rate is not yet well represented by $\mu$, and is expected to be smaller. This leads to a systematic undershoot of the maximal value of $m_r$ in the model as compared to simulations. This then also leads to systematically higher values of cloud mass in the rainfall regime in the model.
On the other hand, choosing the transition time to be at the peak of $m_r$, $\tau_r=\tau_{peak}$ as in the main text, will prolong the mass conserving regime beyond its applicability, leading to an overshoot in the value of $m_r$ at the peak and an undershoot in the cloud mass as compared to simulations. To capture the transitional temporal regime thus probably requires modeling a time dependent sedimentation rate $\mu(t)$, which smoothly increases from zero to the final value of $\mu$. However, this goes beyond the scope of the present work, given that our simple modeling approach already captures the dynamics reasonably well and that our bin microphysics model is itself highly idealized.

\bibliography{refs}

\end{document}

%% file: shortcuts.tex
\newcommand{\beqn}{\begin{equation}}
\newcommand{\eeqn}{\end{equation}}
\newcommand{\beqa}{\begin{eqnarray}}
\newcommand{\eeqa}{\end{eqnarray}}
\newcommand{\beqanonum}{\begin{eqnarray*}}
\newcommand{\eeqanonum}{\end{eqnarray*}}
\newcommand{\beqnonum}{\begin{equation*}}
\newcommand{\eeqnonum}{\end{equation*}}

\newcommand{\bbf}{\begin{bf}}
\newcommand{\ebf}{\end{bf}}
\newcommand{\eqnref}[1]{Eq. (\ref{#1})}

\newcommand{\kg}{\ensuremath{\mathrm{kg}}}
\newcommand{\meter}{\ensuremath{\mathrm{m}}}

\newcommand{\mm}{\ensuremath{\mathrm{mm}}}

\newcommand{\micron}{\ensuremath{\mu\mathrm{m}}}

